\documentclass{article}

\usepackage{arxiv}

\usepackage{arxiv}
\usepackage{slashed}
\usepackage[utf8]{inputenc} 
\usepackage{fontenc,amssymb }
\usepackage{hyperref}       
\usepackage{url}            
\usepackage{booktabs}       
\usepackage{amsfonts}       
\usepackage{nicefrac}       
\usepackage{microtype}      
\usepackage{lipsum}
\usepackage{physics}
\usepackage{graphicx}
\usepackage{mathtools}
\usepackage{arxiv}
\usepackage{slashed}
\usepackage[utf8]{inputenc} 
\usepackage{fontenc,amssymb }
\usepackage{hyperref}       
\usepackage{url}            
\usepackage{booktabs}       
\usepackage{amsfonts}       
\usepackage{nicefrac}       
\usepackage{float}

\usepackage{microtype}      
\usepackage{lipsum}
\usepackage{physics}
\usepackage{graphicx}
\usepackage{mathtools}
\usepackage[utf8]{inputenc} 
\usepackage[T1]{fontenc}    
\usepackage{hyperref}       
\usepackage{url}            
\usepackage{booktabs}       
\usepackage{amsfonts}       
\usepackage{nicefrac}       
\usepackage{microtype}      
\usepackage{lipsum}
\usepackage{graphicx}
\graphicspath{ {./images/} }
\usepackage{authblk}
\usepackage[square,numbers]{natbib}

\newcommand{\T}{\mathcal{T}}

\title{Accelerated paths and Unruh effect: finite time detector response  in (Anti) de Sitter spacetime and Huygen's Principle.
}

\author[a,b,c]{Shahnewaz Ahmed,}
\author[d,e,f]{Mir Mehedi Faruk,}
\author[g]{Muktadir Rahman}

\affil[a]{School of Data and Sciences, BRAC University, 66 Mohakhali, Dhaka 1212. Bangladesh}
\affil[b]{Perimeter Institute for Theoretical Physics,
Waterloo, ON N2L 2Y5, Canada}
\affil[c]{Department of Physics and Astronomy, University of Waterloo, Waterloo, Ontario N2L 3G1, Canada}
\affil[d]{Department of Physics, McGill University,
Montreal, Quebec, H3A 2T8, Canada}
\affil[e]{Institute for Theoretical Physics, University of Amsterdam, Science Park 904, 1090 GL Amsterdam, The Netherlands.}
\affil[f]{Delta Institute for Theoretical Physics, Science Park 904, PO Box 94485, 1090 GL Amsterdam, The Netherlands.}
\affil[g]{Department of Physics,
University of Nevada, Reno
1664 N Virginia St, Reno, NV 89557.}
\usepackage{xcolor}
\begin{document}
\maketitle
\begin{abstract}

We study the finite time response of an Unruh-DeWitt particle detector 
described by a qubit (two-level system) moving with uniform constant acceleration in maximally symmetric spacetimes. The  
$D$ dimensional massless fermionic response function in
de Sitter (dS) background is found to be identical to that of a detector linearly coupled to a massless scalar field in $2D$ dimensional dS background. Furthermore, we visit the status of Huygen's principle 
in the Unruh radiation observed by the detector. \\
\end{abstract}

\large
A uniformly accelerating  observer of constant  acceleration $a$ moving in  Minkowski or maximally
symmetric 
spacetime
sees the vacuum for an
inertial observer as a thermal state of temperature $T=\frac{\omega}{2\pi}$. Here,
$\omega$ is given by\cite{deser1999mapping, deser1997class, deser1998equivalence, padmanabhan2003cosmological, padmanabhan2005gravity,Jennings:2010vk},
\begin{equation}
\omega =  \left\{
        \begin{array}{ll}
            \sqrt{a^2+k^2}, & \quad \text{dS}, \Lambda>0 \\
            \sqrt{a^2-k^2} & \quad \text{AdS}, \Lambda<0
            \\
                a & \quad \textrm{Minkowski}, \Lambda=0
        \end{array}
    \right.
\end{equation}
Here the cosmological constant $\Lambda$ 
is related to $D$ dimension
spacetime curvature, 
$k$ 
 through 
$|\Lambda|=\frac{k^2}{2}(D-2)(D-3).$
In recent times,
Unruh radiation
and its close
analogue 
Hawking radiation has been extensively studied with the tool
of the 
Unruh-Dewitt (UDW)
particle detector. 
The UDW detector  has 
applications
connecting other branches of physics including but not limited to 
understanding
harvesting entanglement\cite{salton2015acceleration, ng2018new, tjoa2021entanglement, maeso2022entanglement}, QCD\cite{caceres2010quantum}, complexity\cite{blommaert2021unruh}, cosmology\cite{conroy2022unruh, martin2012cosmological,Blasco:2015eya}, as well as in application-oriented research directions such as
condensed matter systems\cite{yan2022effect,Bhardwaj:2022sfw} like anyons\cite{ohya2017emergent}
and
constructing
heat engines\cite{Arias:2017kos,ottoEngine} using UDW detector.
The 
accelerated
UDW  detector shows a response when coupled to matter field.
In one of the pioneering work on the topics related to detector physics was by Takagi\cite{takagi1986vacuum} where interesting features of
detector response  
function were elaborated.
A complete story on fermionic response
function to accelerated detectors in flat spacetime
was developed recently in \cite{louko2016unruh}.
It was noted in that \cite{louko2016unruh} in $D$-dimensional Minkowski spacetime,
 the response of the accelerated UDW detector coupled 
to massless Dirac field proportional to that
of a detector linearly coupled to a massless scalar field in $2D$ dimension. 
This observation was quite interesting as it helped us measure the Unruh radiation observed by the detector when coupled to the fermionic matter field. But it is a natural question to ask if a similar 
conclusion also arises
when the fermionic field is coupled to curved spacetime.
In our previous article\cite{ahmed2021accelerated} we explained how the same mechanism works in AdS background instead.
Another significant observation made by several authors\cite{Sriramkumar:2002nt,Ooguri:1985nv} is the apparent statistics inversion
in the Unruh radiation in odd dimensional spacetime.
In odd-dimensional spacetime, we notice that the thermal radiation measured by a linear UDW particle detector coming from scalar field can maintain an anti periodic relation.   
Our previous article explained how non-linearity affects the statistics inversion in AdS spacetime.
Of course, when the curvature of the spacetime was approaching zero, we reproduced the  known results of  detector response in flat spacetime\cite{Jennings:2010vk,ahmed2021accelerated} but the similar setup in the dS background still needs to be  discussed. In this article, we analyze
the effects of non-linearity and develop the
calculation of the fermionic response function in the dS background. In our earlier work and many other studies relating to the UDW detector, we investigated the response function where the detector is "turned on" for an infinite amount of time \cite{ Jennings:2010vk,takagi1986vacuum,  ahmed2021accelerated}.
In practice it is impossible to turn on the detector for infinite time\cite{ottoEngine}; therefore,
the finite time response has been rigorously investigated
in the context of flat space time \cite{Sriramkumar:1994pb}.
We start our manuscript by analysing the finite time response of accelerated an UDW detector in AdS spacetime coupled to real scalar fields in a non-linear way.  In section 2, we first elaborate on the dS case for the real scalar fields and then finish the calculation for fermionic fields.
We prove here the response function of the uniformly accelerated UDW detector coupled to a massless Dirac field in dS spacetime of dimension is equivalent to the response function of the detector linearly coupled to a massless scalar field in 2D dimensional dS spacetime.
 We have generalised the result in the case of non trivial gravitational background AdS spacetime in part-1\cite{ahmed2021accelerated} and dS spacetime in this article.
Finally, we summarise the cases when the Huygens principle is maintained or violated by the Unruh radiation observed by the accelerated detectors in maximally symmetric spacetime. We point out that fermionic Unruh radiation respects the Huygens principle in any dimension, unlike scalar theory. We explain here the reason behind. \footnote{In previous study, (see   ref.\cite{Ooguri:1985nv}),  it was claimed that the Huygens principle is violated (satisfied) in even (odd) dimensions for fermionic Unruh radiation. We show here that it is not the case.} Throughout the whole article, we chose $\hbar = 1$, $c=1$ and Boltzmann constant $k_B = 1$ in our calculation.
\section{Finite time response of UDW detector: Scalar field}
	We first consider  a real scalar field $\Phi$ in $D$ dimensional (A)dS spacetime
which is conformally coupled to gravitational background. 
We are
are considering the AdS metric in Poincare coordinates but 
we can ofcourse choose any other coordinates system such as global coordinates.
Similarly we will follow the flat slicing for De Sitter background.
The AdS metric in Poincare coordinate,
\begin{eqnarray}
ds^{2}=\displaystyle \frac{1}{k^{2}z^{2}} (dt^{2}-dx_{1}^{2}-dx_{2}^{2} - ... -dx_{D-2}^{2} - dz^2).\label{metric1}
\end{eqnarray}
The dS metric in flat slicing is written as
\begin{eqnarray}
ds^{2}=\displaystyle \frac{1}{k^{2}\eta^{2}} (d\eta^{2}-dx_{1}^{2}-dx_{2}^{2} - ... -dx_{D-2}^{2} - dx_{D-1}^{2}).\label{metric2}
\end{eqnarray}
Depending upon what we would like to have as our
 our gravitational background 
 we pick AdS or dS we choose
eq.(\ref{metric1}) or eq. (\ref{metric2}). 
The  total action of our  system of interest is,\\
\begin{eqnarray}
&&S=S_0+S_{int}+S_{detector},
 \label{kukumia}
\end{eqnarray}
The matter field action is simply-
\begin{eqnarray}
S_0=\frac{1}{2}\int d^Dx \sqrt{|g|}\left(g^{\mu\nu}\triangledown_\mu \Phi
\triangledown_\nu \Phi+\zeta R\Phi^2\right).
\end{eqnarray}
When
scalars are conformally coupled to gravity one can specify\cite{Weinberg72},
\begin{eqnarray}
\zeta= \frac{D-2}{4(D-1)}. \label{zeta}
\end{eqnarray}
The interaction part of the Hamiltonian is simply,
\begin{eqnarray}
{H}_I(\tau) = \lambda \chi_{\mathcal{T}}(\hat{\sigma}^-(\tau) + \hat{\sigma}^+(\tau)){\Phi}(z(\tau)),\label{kukumia}
\end{eqnarray}
where, $\lambda$ is the strength of coupling, $\chi_{\mathcal{T}}$ is a switching function which controls the time of interaction with the field, and $n$ is any positive integer. 
${\mathcal{T}}$ represents how long the detector is on. We are going to refer it as switching time.
It is well known that any sudden jump in the switching function may cause divergence\cite{ottoEngine} for
 finite time interaction.  Therefore we choose a  Lorentzian switching function (a smooth function),
 \begin{eqnarray}
 \chi_{\mathcal{T}}(\tau) = \frac{(\mathcal{T}/2)^2}{\tau^2 + (\mathcal{T}/2)^2}\label{33}
 \end{eqnarray}
One can simply set ${\mathcal{T}}\rightarrow \infty$, or $\chi_{\mathcal{T}}=1$ in order to obtain the usual detector response (where it is assumed that the detector can interact with the matter fields for infinite time).    The detector is thought  as a two-level quantum system defined along a worldline $x(\tau)$. The detector  Hamiltonian is,
    \begin{equation}\label{HD}
        \hat{H}_D = \frac{\Omega}{2}\left(\hat{\sigma}^+\hat{\sigma}^- -\hat{\sigma}^-\hat{\sigma}^+\right),
    \end{equation}
    We are thinking the UDW detector as two level system. There are two states   
     $\ket{g}$ and $\ket{e}$ and 
    $\Omega$ is the energy gap between these
    two states.
    The
     $\hat{\sigma}^+$ and $\hat{\sigma}^-$ are the well known $SU(2)$ ladder operators.
     \begin{eqnarray}
&&     \ket{e} = \hat{\sigma}^+\ket{g},\\ &&\hat{\sigma}^-\ket{g} = 0
     \end{eqnarray}
     From this Hamiltonian we can easily see the ground and excited states of the detector are $\ket{g}$ and $\ket{e}$, respectively.
     \begin{eqnarray}
  &&     \hat{H}_D\ket{e}=\frac{\Omega}{2}\ket{e}\\
&&  \hat{H}_D\ket{g}=-\frac{\Omega}{2}\ket{g}
     \end{eqnarray}
     Point to note that $\hat{H}_D$ 
     generates time translations with respect to the detector's proper time $\tau$.
     We are assuming that the detector
follows a timelike trajectory $x(\tau)$ 
which
parametrized by proper time $\tau$ in 
$D$ dimensional
 spacetime.
Any
 real scalar quantum field
 ${\Phi}(x)$ can be written as a mode expansion,
    \begin{equation}
        {\Phi}(x) = \int d^{l}k \left(f_{k}(x) \hat{b}_{k}+f_{k}^*(x) \hat{b}^\dagger_{k}\right),
    \end{equation}
    where $\{u_{ k}(x)\}$ is assumed to be a normalized basis of solutions to {the Klein-Gordon equation}. 
    The functional form is fixed 
    from the gravitational background.
    the annihilation and creation operators maintaining the usual commutation relations
    are 
     $\hat{b}_{ k},\hat{b}_{ k}^\dagger$, respectively. In principle the creation/annihilation operators help can be used  construct a Hilbert space representation for the quantum field, defined in terms of the vacuum state $\ket{0}$.
The most interesting quantity to us is
the probability amplitude related  to the transition from the initial state $\ket{g,0}$ to a state $\ket{e,\varphi}$. Here $\ket{\varphi}$ denotes an arbitrary final state of the field.
 The amplitude can be found following\cite{aBurbano:2020fue}, 
   \begin{align}\label{amplitudephi}
        &\mathcal{A}_{g\rightarrow e}(\varphi)=\bra{e,\varphi}{U}_I\ket{g,0} = \sum_{n=0}^\infty \bra{e,\varphi}{U}_I^{(n)}\ket{g,0}\\=& \sum_{n\textrm{ odd}}\lambda^{n}\frac{(-i)^{n}}{n!}\int \dd{\tau_1}\cdots\dd{\tau_{n}} \chi(\tau_1)\cdots\chi(\tau_{n}) \bra{\varphi}\mathcal{T}\Big(\hat{\phi}(\tau_1)\cdots\hat{\phi}(\tau_{n})\Big)\ket{0}e^{i\Omega(\tau_1-\tau_2+\dots+ \tau_{n})},\nonumber
    \end{align}
    Here ${U}_I$
 the time evolution operator is given in the usual way,
    \begin{eqnarray}
        {U}_I &= \mathcal{T}\exp\left(-i \int \dd \tau {H}_I(\tau)\right)
        = \sum_{n=0}^\infty \underbrace{\frac{(-i)^n}{n!}\int \dd\tau_1\cdots\dd\tau_n \mathcal{T}\Big({H}_I(\tau_1)\cdots{H}_I(\tau_n)\Big)}_{\displaystyle{{U}_I^{(n)}}}=\sum_{n=0}^\infty {U}_I^{(n)}\nonumber \\
    \end{eqnarray}
One can determine the transition probability to arbitrary order by tracing over the field final states $\ket{\varphi}$, 
    \begin{equation}
        \begin{aligned}
        \mathcal{P}_{g\rightarrow e} &= \int \textrm{D}\varphi\, \abs{\mathcal{A}_{g\rightarrow e}(\varphi)}^2\\ &= \sum_{n,m\textrm{ odd}}\lambda^{n+m} \frac{ (-i)^{n-m}}{n!m!}\int \dd{\tau_1'}\dots\dd{\tau_{m}'} \dd{\tau_1}\cdots\dd{\tau_{n}} \chi(\tau_1')\cdots\chi(\tau_{m}') \chi(\tau_1)\cdots\chi(\tau_{n})\\&\hphantom{{}={}}\times \bra{0}\mathcal{T}\Big({\Phi}(\tau_1')\cdots{\Phi}(\tau_{m}')\Big)^\dagger\mathcal{T}\Big({\Phi}(\tau_1)\cdot
        {\Phi}(\tau_{n})\Big)\ket{0}e^{-i
        \Omega(\tau_1'-\dots+ \tau_{m}')}e^{i\Omega(\tau_1-\dots+ \tau_{n})},\label{bigb}
        \end{aligned}
    \end{equation}
In this series  the lowest order term is of second order in the coupling constant $\lambda$. It is expressed as,
    \begin{equation}
\mathcal{P}_{g\rightarrow e}^{(2)} = \lambda^2 \int \dd\tau \dd\tau' \chi(\tau)\chi(\tau') \bra{0}{\Phi}(\tau){\Phi}(\tau')\ket{0} e^{i\Omega(\tau-\tau')}
=\lambda^2 \int \dd\tau \dd\tau' \chi(\tau)\chi(\tau')  
W^{(2)}_{\rm_{D}}(x(\tau),x'(\tau'))
e^{i\Omega(\tau-\tau')}\label{nnnn}
.
    \end{equation}
Here,  
$W^{(2)}_{\rm_{D}}(x(\tau),x'(\tau'))$
is the $D$ dimensional two point correlator (Wightman function).
The exact functional form of the Wightman 
function again depends upon the background gravity.

We can consider more general interaction Hamiltonian,
	\begin{equation}
	\mathcal{H}_\text{I}=\lambda\;
	\chi_\mathcal{T}(\tau)m(\tau)\;\mathcal{O}_{\Phi}[x(\tau)]\;\label{nnh},
	\end{equation}
 Here, $m(\tau)$ the monopole operator, 
	\begin{equation}\label{eq:Monpl}
	m(\tau)=e^{i\Omega\tau}\ket{e}\bra{g}+e^{-i\Omega\tau}\ket{g}\bra{e}=\begin{pmatrix}
	0&e^{+i\Omega\tau}\\
	e^{-i\Omega\tau}&0\\
	\end{pmatrix}\: .
	\end{equation}
This actually 
takes the ground state of the detector to the excited state and vice versa. 
In other words we can
 physically 
 describe the procedure
 as "a click"
 in response to the presence of the field.
Now the the operator $\mathcal{O}_{\Phi}$ 
outlines how
the matter  field is coupled to the detector. 
Instead of usual linear coupling we take a more  general coupling\footnote{normal ordering is assumed},
	\begin{equation}\label{eq:SCoup}
	\mathcal{O}_\Phi[x(\tau)]=\Phi^n[x(\tau)]\; 
	\end{equation} 
	If we use the interaction Hamiltonian \eqref{eq:SCoup} then we re-express eq. \eqref{nnnn},
	\begin{eqnarray}
	\mathcal{P}_{g\rightarrow e}^{(2)} =
	\lambda^2 \int \dd\tau \dd\tau' \chi(\tau)\chi(\tau')  
W^{(2n)}_{\rm_{D}}(x(\tau),x'(\tau'))
e^{i\Omega(\tau-\tau')}\label{sss}
	\end{eqnarray}
Here, $W^{(2n)}_{\rm_{D}}(\tau-\tau')=
\bra{0}:\Phi^n(x(\tau)):
:\Phi^n(x(\tau')):\ket{0}$ is the $2n$-point correlator. The detector response function of the UDW detector   is directly proportional to the probability for
the detector to transition from ground state to excited state.
Using  Lorentzian switching function eq. \eqref{33} 
the response function $\mathcal{F}^{(2n)}(\Omega, \mathcal{T})$ will be,
\begin{equation}
 \mathcal{F}^{(n)}(\Omega, \mathcal{T}) = \frac{\pi\mathcal{T}^3}{4}\int^\infty_{-\infty}\dd{(\Delta \tau)}\frac{{W^{(n)}_{\rm_{D}}}(\Delta\tau)}{\Delta\tau^2 + \mathcal{T}^2}e^{-i\Omega\Delta\tau}  \label{chapo}
\end{equation}
 The $2n$-point function $W_D^{(2n)}\left({ x},
{ x'}\right)$ is related to the the Wightman function in the following interesting but simple way by Wick's theorem \cite{das},
\begin{equation}
{W}^{(2n)}_{\rm _{D}}\left({ x}, { x'}\right)
= \left(n!\right)\, \left(W_{\rm_{D}}^{(2)}\left({ x},
{ x'}\right)\right)^{n}.\label{eqn:2nptfnmv}
\end{equation}

\subsection{Finite time response of scalar fields: AdS spacetime}
For conformally coupled scalars the Wightman function in $D>2$ dimensional AdS
spcatime  can be  obtained
in the following form with suitable boundary condition\cite{Jennings:2010vk},
\begin{eqnarray}
W^{(2)}_{\rm AdS_{D}}(x,x')=
\bra{0}\Phi(x(\tau))
\Phi(x(\tau'))\ket{0}={\cal C}_{D} \bigg(\frac{1}{(v-1)^{D/2-1}}-\frac{1}{(v+1)^{D/2-1}}\bigg).\label{twotwo}
\end{eqnarray}
Here, $\nu$ is the conformal invariant defined as,
\begin{eqnarray}
 v = \frac{z^{2}+z^{\prime 2}+(\mathbf{x}-\mathbf{x}')^{2}-(t-t'-i\epsilon)^{2}}{2zz'}\label{nu}.
\end{eqnarray} ${\cal C}_{D}$ is a constant,
\begin{eqnarray}
 {\cal C}_{D}= \frac{k^{D-2}\Gamma(D/2-1)}{2(2\pi)^{D/2}}, \label{CDCD}
\end{eqnarray}
In this article we mainly focus on 
 Unruh effect is a widely studied phenomena which basically states an accelerating observer (with constant linear acceleration $a$) will observe a thermal bath with temperature $T$. If the observer were in flat spacetime,
 the
  temperature is given by the following formula
  $T=\frac{\hbar a}{2\pi c k_B}$.
In the usual literature\cite{Carroll:2004st}  (studies mostly done in flat spacetime) the path which was chosen for the accelerating observer had linear uniform acceleration $a$.
The accelerating observer (detector) can take a circular path with constant velocity $v$, will end up having constant acceleration $a$.  Of course the resultant radiation (detector response) due to this type of non-linear motion will  not be quite thermal radiation  as the correlators will not obey the KMS  relation  \cite{takagiReview}. We are interested in those accelerated  paths 
which corresponds to Wightman function maintaining valid KMS relation.

\subsubsection{Super critical accelerated paths in AdS}
We are  considering the supercritical paths ($a>k$) as only these paths results in non zero response function for the detectors\cite{Jennings:2010vk} in uniform linear acceleration.
In our recent article\cite{ahmed2021accelerated}
we successfully showed that using GEMS (Global Embedding Minkowski Spacetimes) approach
that one can construct a path
with constant acceleration
by considering the path as an intersection between a  flat plane of dimension $M$ 
and $D$ dimensional AdS
hypersurface embedded in $D+1$ dimensional flat spacetime. Here, $M(M<D+1)$.
We also proved that for any uniform linear supercritical trajectories  would have the same conformal invariant $v$ as a function of proper time.  
\begin{eqnarray}
v(\tau,\tau')=\frac{a^2}{\omega^2}- \frac{k^2}{\omega^2}\cosh(\omega(\tau-\tau')-i\epsilon)\label{path222}.
\end{eqnarray}
Here $\omega = \sqrt{a^2-k^2}$.
The example of super critical
path (with constant linear acceleration) in  $z-t$ plane is given in ref. \cite{ahmed2021accelerated},
\begin{equation}
t(\tau)=\frac{a}{\omega}e^{\omega \tau} \;\;,\;\;
z(\tau)= e^{\omega\tau} \;\;,\;\;
x^{1} =x^{2}=x^{3}=\ldots=x^{D-2}=0. \label{eq12}
\end{equation}
Another supercritical  path with constant acceleration $a$ in the $x_1-t$ plane is given by 
the following manner\cite{ahmed2021accelerated},
\begin{eqnarray}
z(\tau)=z_0\;\;,\;\;
x^{1}(\tau)=\frac{z_0k}{\omega} \cosh(\omega \tau) \;\;,\;\;
t(\tau)=\frac{z_0k}{\omega}
\sinh(\omega \tau)
\;\;,\;\;
x^{2}=x^{3}=\ldots=x^{D-2}=0.
\label{eqnxt}
\end{eqnarray}
$z_0$ is a constant and $\tau$ is the proper time.
We could  also define the \eqref{eqnxt} path in the $x^i-t$ direction.
However we have already showed in ref.\cite{ahmed2021accelerated}
that how all uniform accelerating paths with constant acceleration are related by AdS isometries.
Any supercritical path will result in eq. \eqref{path222}. Therefore,  
following eq. \eqref{twotwo}, 
the two point function 
for uniform acceleration (in any supercrtical path) becomes,
\begin{eqnarray}
     G_{\rm AdS_{D}}(\Delta\tau) &=& \frac{\omega^{D-2}\Gamma(\frac{D}{2}-1)}{(4\pi)^{\frac{D}{2}}}\bigg(\frac{1}{i^{D-2}\sinh^{D-2}(\frac{\omega\Delta\tau}{2}-i\epsilon)}
    \nonumber \\
    & \ & \ \ \ -\frac{1}{(\sinh(A+(\frac{\omega\Delta\tau}{2}-i\epsilon)))^{\frac{D}{2}-1}(\sinh(A-(\frac{\omega\Delta\tau}{2}-i\epsilon)))^{\frac{D}{2}-1}}\bigg).\nonumber\\
\end{eqnarray}
Here, $\sinh{A} = \omega/k$.
\begin{figure}[H]
 \begin{center}
\includegraphics[scale=0.8]{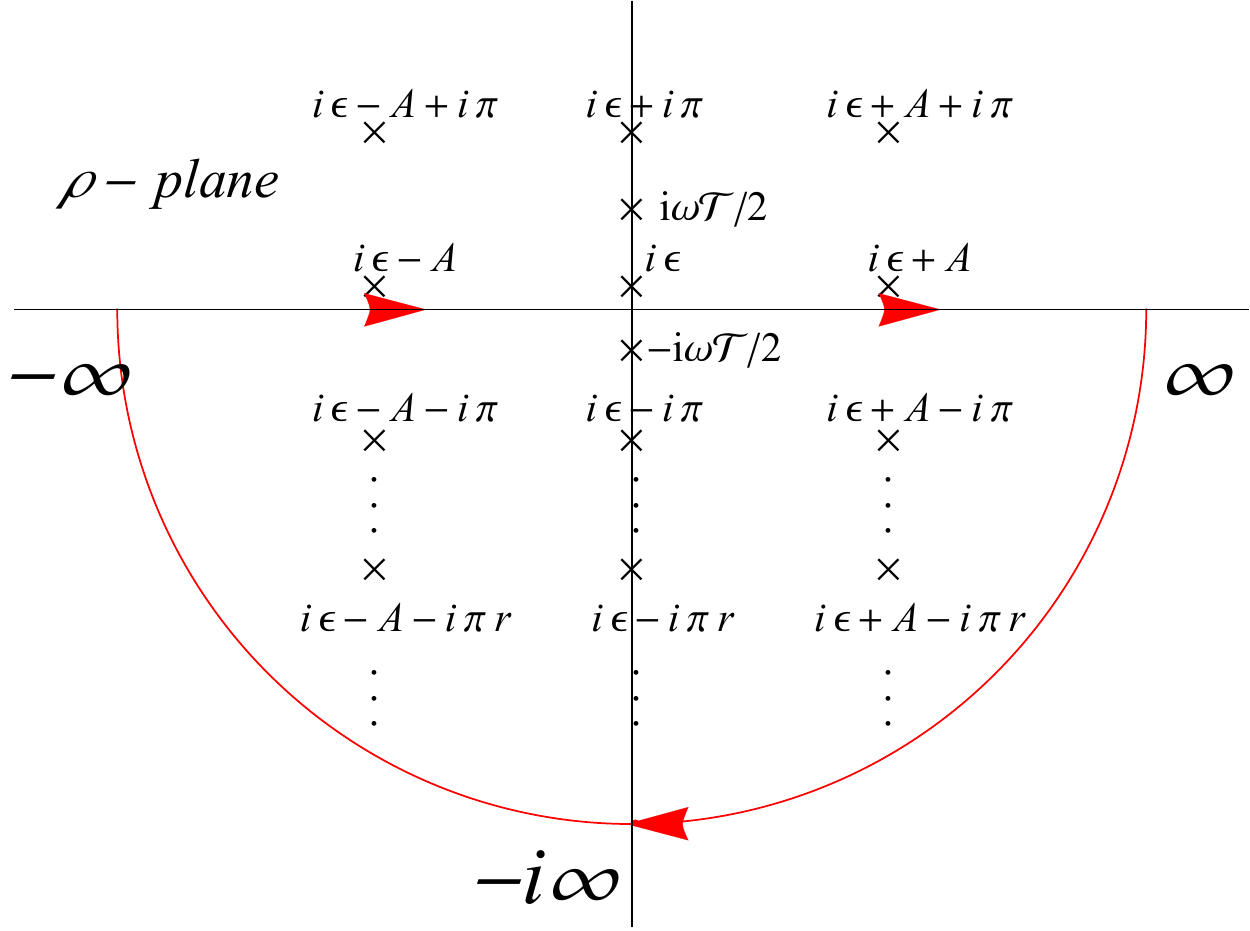}
\hfill
\end{center}
\caption{\label{fig:supc_pole} Contour for evaluating $I_{D, n, \alpha}$.}
\end{figure}
In our previous article we demonstrated the following relation about the two-point correlator\cite{ahmed2021accelerated}, 
\begin{eqnarray}
    {W}^{(2n)}_{AdS_{D}}(\Delta\tau+\frac{2\pi\dot{i}}{\omega}) 
      = (-1)^{nD} {W}^{(2n)}_{AdS_{D}}(\Delta\tau). \label{KMScondition}
\end{eqnarray}
\begin{figure}[H]
\centering 
{\large \textbf{\qquad \qquad \underline{\underline{$F^{(n)}_{{\rm AdS}_4}(\Omega, \T)$ vs. $\Omega$}}}}\\
\hfill 
\includegraphics[scale=0.265]{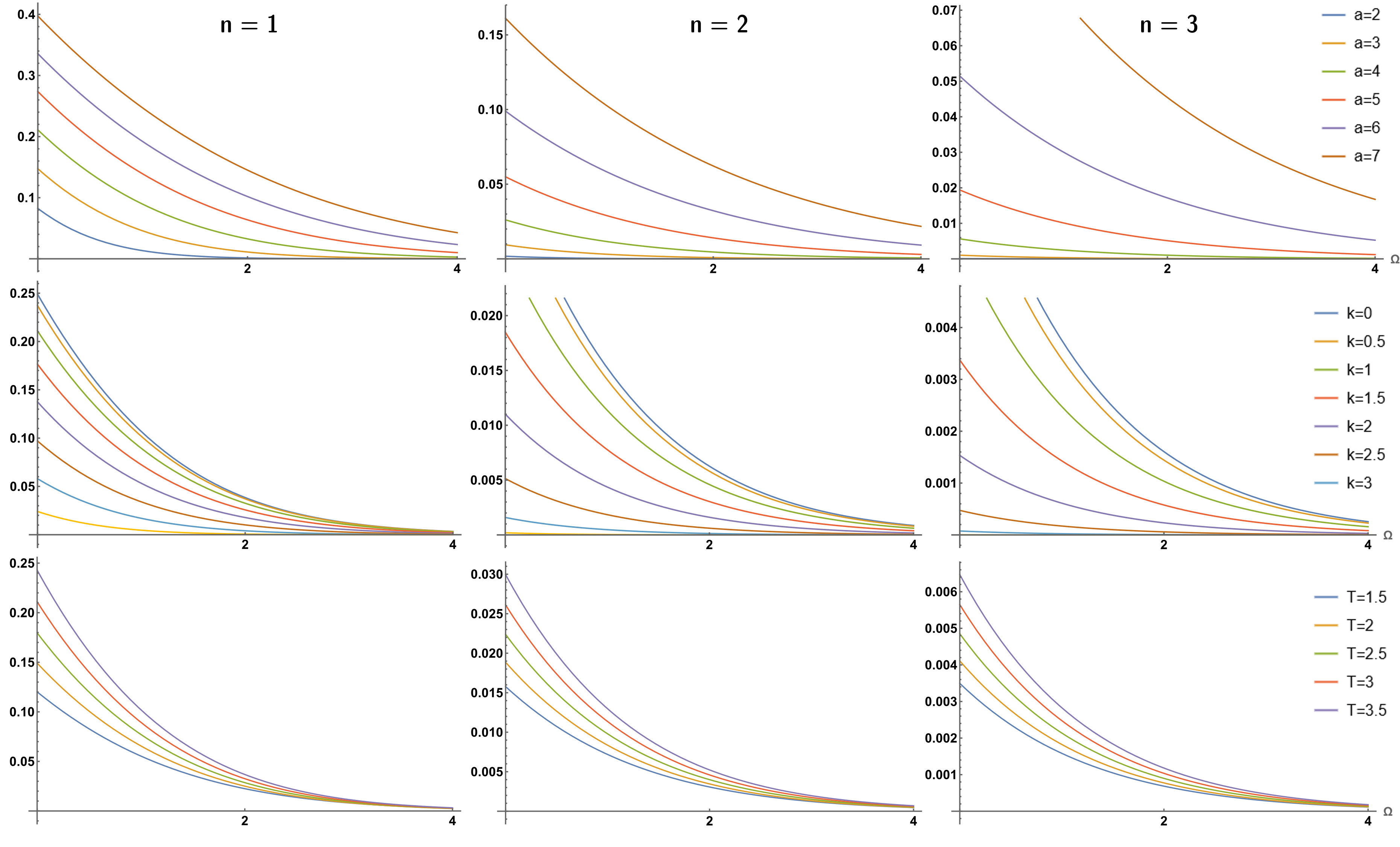}
\hfill
\caption{\label{fig:ads_scalar_vs_omega} Plot of the finite-time response of the UDW particle detector in AdS Space-time against energy $\Omega$. (From left to right) 1st, 2nd and 3rd columns show the plots for different values of $n$, with  $n=1$, $n=2$ and $n=3$. Each row (from top to bottom) shows the variation of the response function with changing $a$ (fixed $k=1,$ $\T=3$), changing $k$ (fixed $a=4$, $\T=3$) and changing $\T$ (fixed $a=4,$ $k=1$) respectively. }
\end{figure}

Therefore $\omega=\sqrt{a^2-k^2}$
is the temperature\cite{ahmed2021accelerated}.
We rewrite the $2n$-point function $G^{(n)}_{{\rm AdS}_D}$ in the following manner \cite{ahmed2021accelerated}, 
\begin{equation}
    G_{\rm AdS_{D}}^{(n)}(\Delta\tau)=(n!) {\cal C}_{D}^{n} \bigg(
\frac{\omega}{\sqrt{2}k}\bigg)^{n(D-2)} \sum^{n}_{\alpha = 0} \binom{n}{\alpha} \frac{(-1)^{\alpha}}{i^p} \
\ \mathcal{G}_{D,n,\alpha} (\rho)
\end{equation}
where,
\begin{eqnarray}    &&\mathcal{G}_{D,n,\alpha} (\rho) = (\sinh(\rho-i\epsilon))^{-p} (\sinh(A + (\rho-i\epsilon)))^{-q} (\sinh(A-(\rho-i\epsilon)))^{-q}\\
&&\rho = \omega \Delta \tau/2 \label{rhooo}\\
&&p = 2(n-\alpha)(D/2-1)\\
&& q = \alpha(D/2 - 1).
\end{eqnarray}\\
So, finally the expression for  the finite time response function in AdS spacetime for our interaction Hamiltonian becomes, 
\begin{eqnarray}
   \mathcal{F}^{(n)}(\Omega, \mathcal{T}) & = \frac{\pi n!\mathcal{T}^3\mathcal{C}_D^n}{4}\bigg(
\frac{\omega}{\sqrt{2}k}\bigg)^{n(D-2)}\int^\infty_{-\infty}\dd{(\Delta \tau)}\frac{e^{-i\Omega\Delta\tau}}{\Delta\tau^2 + \mathcal{T}^2}\sum^{n}_{\alpha = 0} \binom{n}{\alpha} \frac{(-1)^{\alpha}}{i^p}\mathcal{G}_{D,n,\alpha} (\rho)\\
    & = \frac{\pi n!\mathcal{T}^3\mathcal{C}_D^n}{4}\bigg(
\frac{\omega}{\sqrt{2}k}\bigg)^{n(D-2)}\sum^{n}_{\alpha = 0} \binom{n}{\alpha} \frac{(-1)^{\alpha}}{i^p} \left(\frac{\omega}{2}\right) \underbrace{\int^\infty_{-\infty}\dd{\rho}\frac{e^{-i(2\Omega/\omega)\rho}}{\rho^2 + (\omega\mathcal{T}/2)^2}{\mathcal{G}}_{D, n, \alpha}(\rho)}_{\mathcal{F}_{D, n, \alpha}}\\
    & = \frac{\omega\pi n!\mathcal{T}^3\mathcal{C}_D^n}{8}\bigg(
\frac{\omega}{\sqrt{2}k}\bigg)^{n(D-2)}\sum^{n}_{\alpha = 0} \binom{n}{\alpha} \frac{(-1)^{\alpha}}{i^p} {\mathcal{F}_{D, n, \alpha}}\addtocounter{equation}{1}
\end{eqnarray}
where, 
\begin{equation}
\mathcal{F}_{D, n, \alpha} = \int^\infty_{-\infty}\dd{\rho}\frac{e^{-i(2\Omega/\omega)\rho}}{\rho^2 + (\omega\mathcal{T}/2)^2}{\mathcal{G}}_{D, n, \alpha}(\rho).
\end{equation}

\begin{figure}[!h]
\centering 
{\large \textbf{\qquad \qquad \underline{\underline{$F^{(n)}_{{\rm AdS}_4}(\Omega, \T)$ vs. $a$}}}}\\
\hfill 
\includegraphics[scale=0.35]{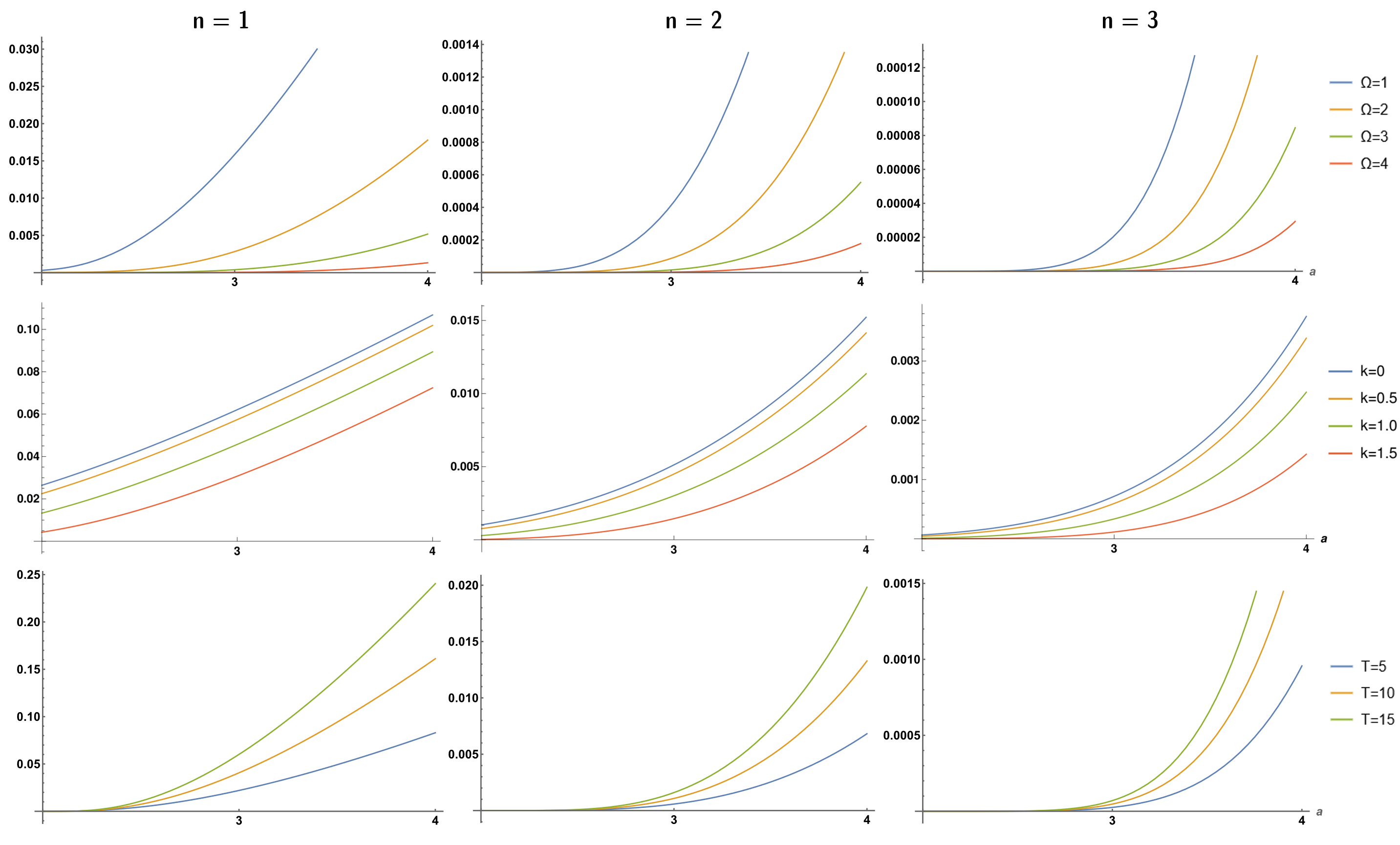}
\hfill
\caption{\label{fig:ads_scalar_vs_a} Plot of the Finite-Time response of the UDW Particle detector in AdS Space-time against acceleration $a$. (From left to right)  1st, 2nd and 3rd columns show the plots for different values of $n$, with  $n=1$, $n=2$ and $n=3$, respectively. Each row (from top to bottom) shows the variation of the response function with changing $\Omega$ (fixed $k=2$, $\T=3$), changing $k$ ($\Omega=1$, $\T=3$) and changing $\T$ ($\Omega=1,$ $k=2$) respectively. }
\end{figure}
Next, we evaluate $\mathcal{F}_{D, n, \alpha}$ by computing the contour integral using semi-circle contour containing the lower half of complex $\rho$ plane (shown in Fig. \ref{fig:supc_pole}). Thus we obtain,
\begin{equation}
\begin{split}
    & \mathcal{F}_{D,n,\alpha}(\mathcal{T})  =  -2\pi i \times \big\{ \text{sum of the residues at } \rho=-i\pi r , \pm A-i\pi r \ (\text{ where } r = 1, 2, ...  ) \  \\
    &  \ \ \ \ \ \ \ \ \ \ \ \ \ \ \  \text{ and } \rho = -i\omega \mathcal{T}/2 \text{ of }  \frac{\mathcal{G}_{D,n,\alpha}(\rho)}{\rho^2 + \left(\omega\mathcal{T}/2\right)^2}   \big\} \\
    & =  -2\pi i \times \Bigg( \lim_{\rho \to -i\omega \mathcal{T}/2} \  \frac{e^{-i(2\Omega/\omega)\rho}}{\sinh^{q}{(A+\rho)}\sinh^{q}{(A-\rho)} \sinh^{p}{(\rho)}\left(\rho -  \left(\frac{i\omega\mathcal{T}}{2}\right) \right)}  + \\
    & \sum^{\infty}_{r=1}
    \bigg[\lim_{\rho \to -i\pi r-A} \frac{\eta(q-1)}{\Gamma(q)} \bigg( \frac{1}{\cosh (\rho+A)} \frac{d}{d\rho} \bigg)^{q-1} \frac{e^{-i(2\Omega/\omega)\rho}}{\cosh{(A+\rho)}  \sinh^q{(A-\rho)} \sinh^{p}{(\rho)} \left(\rho^2 + \left(\frac{\omega\mathcal{T}}{2}\right)^2 \right)} +  \\
    & \ \ \ \ \ \ \ \lim_{\rho \to -i\pi r+A} \frac{\eta(q-1)}{\Gamma(q)} \bigg( \frac{-1}{\cosh (A-\rho)} \frac{d}{d\rho} \bigg)^{q-1} \frac{-e^{-i(2\Omega/\omega)\rho}}{\sinh^q{(A+\rho)} \cosh{(A-\rho)} \sinh^{p} {(\rho)} \left(\rho^2 + \left(\frac{\omega\mathcal{T}}{2}\right)^2 \right)} +\\
    & \ \ \ \ \ \ \ \lim_{\rho \to -i\pi r} \frac{\eta(p-1)}{\Gamma(p)} \bigg( \frac{1}{\cosh \rho} \frac{d}{d\rho} \bigg)^{p-1} \frac{e^{-i(2\Omega/\omega)\rho}}{\sinh^q{(A+\rho)} \sinh^q{(A-\rho)} \cosh {(\rho)} \left(\rho^2 + \left(\frac{\omega\mathcal{T}}{2}\right)^2 \right)} \bigg] \Bigg)
\end{split}
\end{equation}
Here, $\eta(p)$ is Heaviside step function and $\Gamma(p)$ is gamma function. 
 In our previous case we were able to analytically solve the four dimensional response function in AdS when we
chose the infinite switching time, i.e. 
${\mathcal{T}}=\infty$. However it was not possible to calculate the detector response for finite switching time.
We have evaluated the response function numerically
in finite switching time.
Finally we have plotted the response function 
in different ranges of parameter space.
In figure 2, we plot the detector response as a function of energy gap of the two level UdW detector. In the three different columns of figure 2, we have chosen three values of $n$. 
We can see that
when the detector energy gap increases the response goes to zero. In the first row we have fixed the curvature and switching time while varying the energy. We can clearly see the response weakens with greater value of $n$. But if we increase the acceleration  of the detector  the temperature of the radiation also increases. Therefore the first row of Fig 2 manifests response function with respect to $\Omega$ takes greater value when  acceleration is increased. Similar trends are noticed when we look at the other two plots of figure 1. It is very interesting to see as the curvature of AdS spacetime goes to zero 
the detector response rises. Complete opposite trends are noticed in dS spacetime.
But in both cases we have better response function
if we can "turn on" the detector for longer time.\\
\begin{figure}[!h]
\centering 
{\large \textbf{\qquad \qquad \underline{\underline{$F^{(n)}_{{\rm AdS}_4}(\Omega, \T)$ vs. $k$}}}}\\
\hfill\\ 
\includegraphics[scale=0.2]{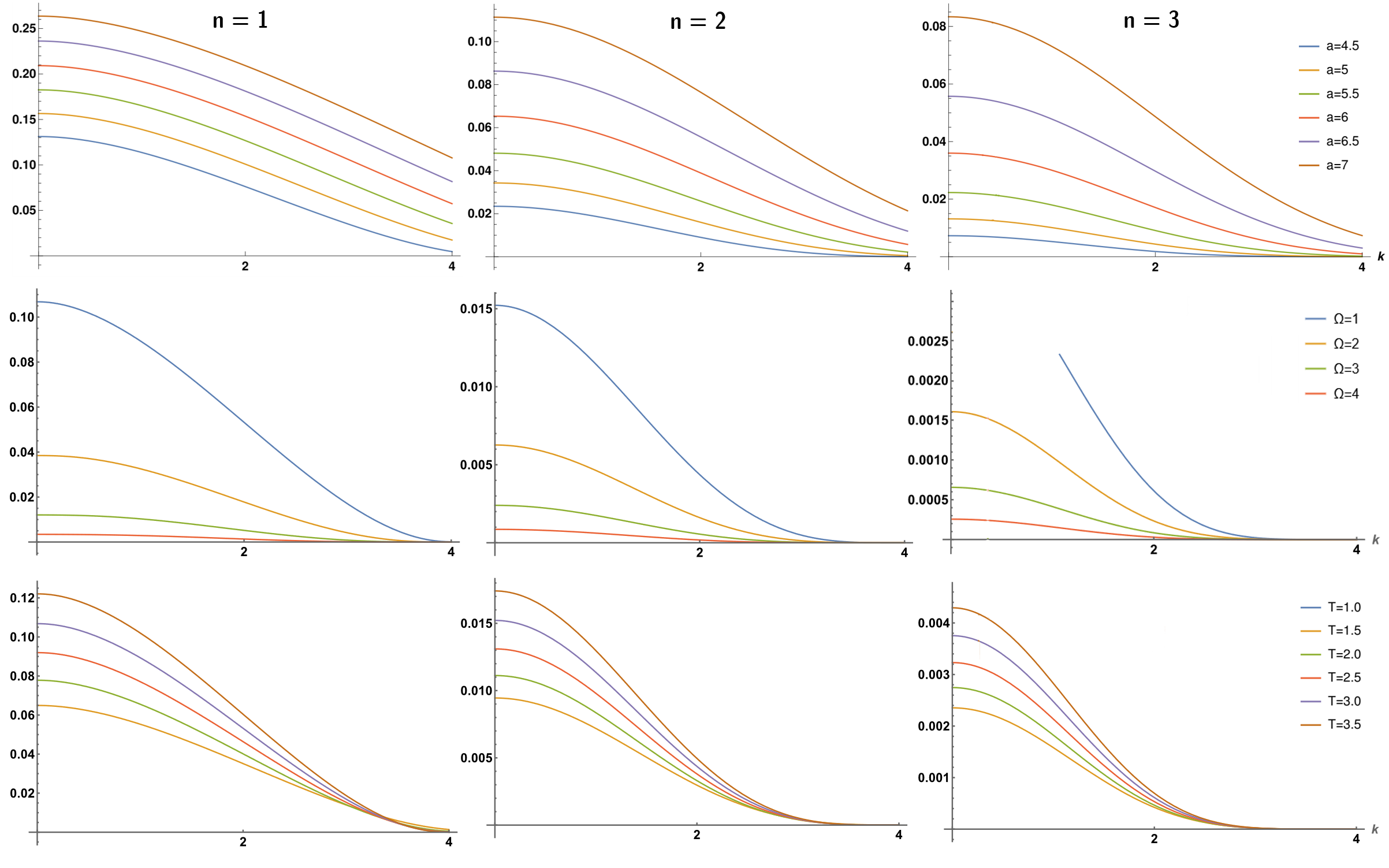}
\hfill
\caption{\label{fig:ads_scalar_vs_k} Plot of the Finite-Time response of the UDW Particle Detector in AdS Space-time against the curvature of AdS ($k$) for massless scalar fields. (From left to right) 1st, 2nd and 3rd columns show the plots for the $n=1$, $n=2$ and $n=3$ coupling to the scalar field. Each row (from top to bottom) shows the variation of the response function with changing $a$ ($\T=3, \Omega=1$), changing $\Omega$ ($a=4, \T=3$) and changing $\T$ ($a=4, \Omega=1$) respectively.}
\end{figure}
As expected from previous discussion if we can accelerate the detectors more and more, we should be able have best response from the detector. We also see it in figure 3.
In the first row of figure 3 we can notice when energy gap rises the response weakens.
Even if the detector is accelerated with higher value it will be difficult to excite the detector if the energy gap is too much.
This problem persists more if we have non-linear coupling between the matter field and the detector. The trend of response function changing with curvature is very interesting. As the curvature increases more and more it becomes problematic to excite the detector. However, we should point out AdS spacetime is a constant curvature spacetime. So when we plot for different values of $k$, we mean that
we are comparing the results the results of response function
for different AdS spacetime 
with different curvatures.
We do not mean that we are analysing the detector spectrum where the gravitational 
background has varying curvature.

\subsection{Scalar field in dS space}
We now
 we use the same setup of two level detector as before but now we consider that the background spacetime is de Sitter spacetime. The metric of de Sitter spacetime can be expressed by so called flat slicing as below-
\begin{eqnarray}
ds^{2}=\displaystyle \frac{1}{k^{2}\eta^{2}} (d\eta^{2}-dx_{1}^{2}-dx_{2}^{2} - ... -dx_{D-1}^{2} ).\label{metric}
\end{eqnarray}
Now, we are going to discuss about the real scalar field $\Phi$ that is conformally coupled to dS gravitational background. We have the same matter field action as well as the same interaction Hamiltonian as in previous section (eq. \eqref{nnh}).
Just as before we need to know the two-point correlator (Wightman function) in order to evaluate the detector response.
The Wightman  function for "Euclidean" vacuum $\ket{0}$ can easily be obtained for conformally coupled real scalar field \cite{Strominger02,takagiReview},
\begin{eqnarray}
W^{(2)}_{\rm dS_{D}}(x,x')=
\bra{0}\Phi(x(\eta))
\Phi(x(\eta'))\ket{0}={\cal K}_{D} v^{1-D/2}\label{two}
\end{eqnarray}
where, 
\begin{eqnarray}
&&{\mathcal{K}}_{D}= \frac{k^{D-2}\Gamma(D/2-1)}{2(2\pi)^{D/2}}. \label{CDCD}
\end{eqnarray}
Here $\nu$ is the conformal invariant.
\begin{eqnarray}
 \nu = \frac{(\vec{x}-\vec{x'})^{2}-(\eta-\eta'-i\epsilon)^{2}}{2\eta \eta'}\label{nu}.
\end{eqnarray}
In order to examine Unruh radiation through the detectors need to move through a constant accelerated path in dS background.
An example of accelerating path in dS spacetime with constant linear acceleration (see Appendix \ref{dS_acc_proof}), we can choose the following,
\begin{equation}
\eta(\tau)= \tau_0 e^{\omega\tau}  \;\;,\;\;
x^{1}(\tau)= \frac{a}{\omega} \tau_0 e^{\omega \tau} \;\;,\;\;
x^{2}=x^{3}=\ldots=x^{D-1}=0, \label{eq12}
\end{equation}
where $\omega = \sqrt{a^2+k^2}$. 
Plugging $\eta$ and $x^{i}$ from eq. (\ref{eq12}) to eq. (\ref{nu}), the conformal invariant $\nu$ takes the following form in this case,
\begin{align}
    \nu &= \frac{(\frac{a}{\omega}\tau_0e^{\omega\tau} - \frac{a}{\omega}\tau_0e^{\omega\tau'})^2 - (\tau_0e^{\omega\tau} - \tau_0e^{\omega\tau'})^2 }{2\tau_0^2e^{\omega\tau}e^{\omega\tau'}}\nonumber\\
    &=\frac{1}{2}\left(\frac{a^2}{\omega^2}-1\right)\left(\frac{e^{\omega\tau} - e^{\omega\tau'}}{e^{\omega(\tau+\tau')/2}}\right)^2\nonumber\\
    &=-\frac{H^2}{2\omega^2}\left(e^{\omega\Delta\tau/2} - e^{-\omega\Delta\tau/2}\right)^2\nonumber\\
    &=-\frac{2H^2}{\omega^2}\sinh^2(\omega\Delta\tau/2)
\end{align}
Following eq. \eqref{two}, 
the two point function for uniformly accelerating paths becomes,
\begin{eqnarray}
     G_{\rm dS_{D}}(\Delta\tau) &=& \frac{\omega^{D-2}\Gamma(\frac{D}{2}-1)}{(4\pi)^{\frac{D}{2}}} \frac{1}{i^{D-2}\sinh^{D-2} (\omega\Delta\tau/2-i\epsilon)}.
\end{eqnarray}
\begin{figure}[!h]
\centering 
{\large \textbf{\qquad \qquad \underline{\underline{$F^{(n)}_{{\rm dS}_4}(\Omega, \mathcal{T})$ vs. $\Omega$}}}}\\
\hfill 
\includegraphics[scale=0.4]{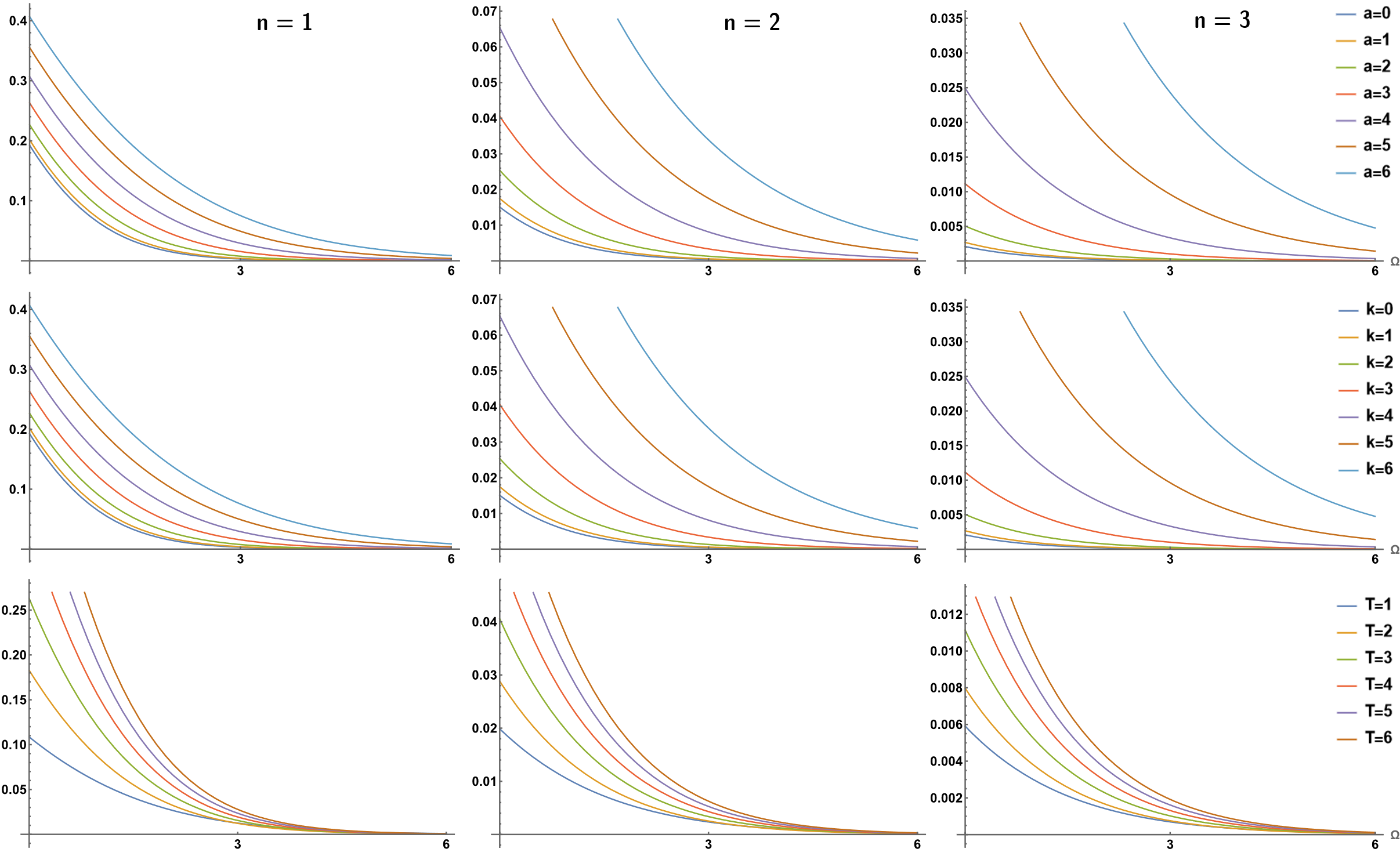}
\hfill
\caption{\label{fig:dS_scalar_response_vs_omega} Plot of the finite-time response of the UDW particle detector in dS space-time against energy $\Omega$. (From left to right) 1st, 2nd and 3rd columns show the plots for different values of $n$, with  $n=1$, $n=2$ and $n=3$. Each row (from top to bottom) shows the variation of the response function with changing $a$ (fixed $k=3,$ $\T=3$), changing $k$ (fixed $a=3$, $\T=3$) and changing $\T$ (fixed $a=3,$ $k=3$) respectively. }
\end{figure}

\begin{figure}[!h]
\centering 
{\large \textbf{\qquad \qquad \underline{\underline{$F^{(n)}_{{\rm dS}_4}(\Omega, \T)$ vs. $a$}}}}\\
\hfill 
\includegraphics[scale=0.39]{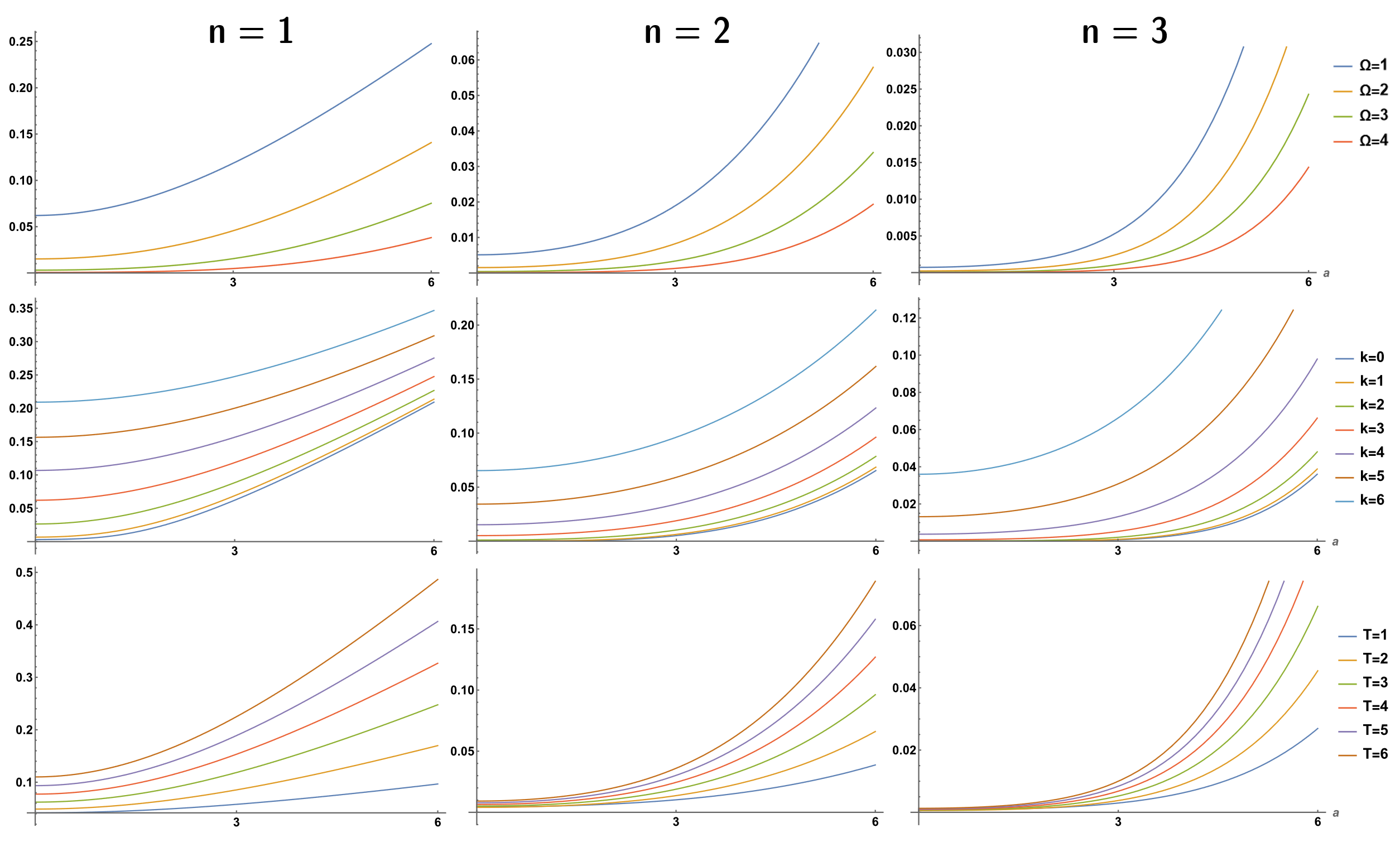}
\hfill
\caption{\label{fig:dS_scalar_response_vs_a} Plot of the Finite-Time response of the UDW Particle detector in dS Space-time against acceleration $a$. (From left to right)  1st, 2nd and 3rd columns show the plots for different values of $n$, with  $n=1$, $n=2$ and $n=3$. Each row (from top to bottom) shows the variation of the response function with changing $\Omega$ (fixed $k=3$, $\T=3$), changing $k$ ($\Omega=1$, $\T=3$) and changing $\T$ ($\Omega=1,$ $k=3$) respectively. }
\end{figure}
We can define the transition probability rate or detector's response function (per unit time)\footnote{In this scenario, we are considering the detector can be switched on for infinite time.}
for interaction Lagrangian \eqref{kukumia} of scalars\cite{Weinberg72},
\begin{eqnarray}
\mathcal{F}^{(n)}_{\rm dS_{D}}=\int_{-\infty}^\infty d\Delta\tau e^{-iE\Delta \tau}
W^{(2n)}_{\rm dS_{D}}
(\Delta\tau).\label{eqn:detrate}
\end{eqnarray}
Here, $W_{\rm dS_{D}}^{(2n)}(\tau-\tau')=
\bra{0}:\Phi^n(x(\tau)):
:\Phi^n(x(\tau')):\ket{0}$ is the $2n$ correlator.
The $2n$-point function $W_{\rm dS_{D}}^{(2n)}$ is related to the the Wightman function in the following way by Wick's theorem \cite{das},
\begin{equation}
W^{(2n)}_{\rm dS_{D}}\left({ x}, { x'}\right)
= \left(n!\right)\, \left(W^{(2)}_{\rm dS_{D}}\left({ x},
{ x'}\right)\right)^n.\label{eqn:2nptfnmv}
\end{equation}
So, the $2n$ correlator becomes,
\begin{eqnarray}
W_{\rm dS_{D}}^{(2n)}(\Delta\tau)= (n!){\cal K}_{D}^{n} \bigg(
\frac{\omega}{\sqrt{2}H}\bigg)^{n(D-2)}\bigg(\frac{1}{i^{D-2}\sinh^{D-2}(\frac{\omega\Delta\tau}{2}-i\epsilon)}\bigg)^n.
\label{eqn:2nptfnmvrind}
\end{eqnarray}

Now the KMS condition can be easily checked using the equation (\ref{eqn:2nptfnmvrind}).
\begin{eqnarray}
    {W}^{(2n)}_{dS_{D}}(\Delta\tau+\frac{2\pi{i}}{\omega}) &=& (n!){\cal K}_{D}^{n} \bigg(\frac{\omega}{\sqrt{2}H}\bigg)^{n(D-2)}\bigg(\frac{1}{i^{D-2}\sinh^{D-2}(\frac{\omega}{2}\left(\Delta\tau+\frac{2\pi{i}}{\omega}\right)-i\epsilon)}\bigg)^n \nonumber\\
    &=&(-1)^{nD}{W}^{(2n)}_{dS_{D}}(\Delta\tau).\label{ss}
\end{eqnarray}
This behavior is similar to the AdS space for nonlinear coupling \cite{jhep} with a major difference with radiation temperature being $\omega=\sqrt{a^2+k^2}$\cite{deser1998equivalence}. We can obtain the Unruh-Dewitt detector response function by taking $\alpha \to 0$ in equation (37) in \cite{jhep}. 
\begin{equation}
   \mathcal{F}_{dS_D}^{(n)} = \Bigg(n! \ {\cal K}_{D}^{n} \bigg(\frac{\omega}{\sqrt{2}H}\bigg)^{n(D-2)}
     \frac{(-1)^{n(D-2)+1}}{i^{n(D-2)}} \  \frac{2}{\omega} I_{D,n}\Bigg)\frac{1}{e^{2\pi E /\omega} - (-1)^{n(D-2)}}. \label{responseFunction}
\end{equation}
where,
\begin{eqnarray}
    I_{D,n} = 2\pi i \times  \frac{1}{\Gamma(n(D-2))} \lim_{\rho \to 0} \bigg( \bigg( \frac{1}{\cosh \rho} \frac{d}{d\rho} \bigg)^{n(D-2)-1} \frac{e^{-i\frac{2E}{\omega}\rho}}{ \cosh {(\rho)} } \bigg).
\end{eqnarray}
\begin{figure}[tbp]
\centering 
{\large \textbf{\qquad \qquad \underline{\underline{$F^{(n)}_{{\rm dS}_4}(\Omega, \T)$ vs. $k$}}}}
\hfill 
\includegraphics[scale=0.39]{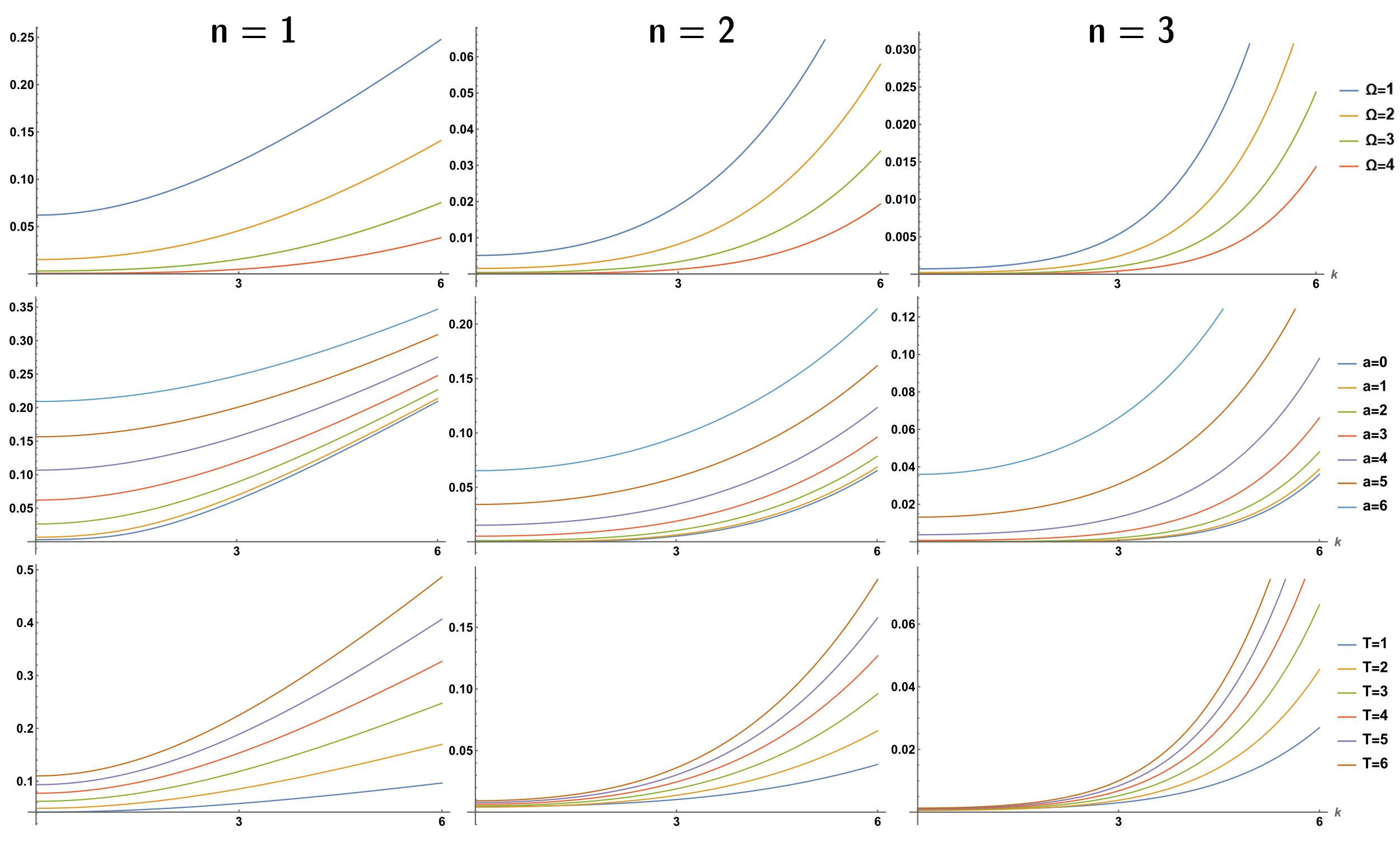}
\hfill
\caption{\label{fig:dS_scalar_response_vs_k} Plot of the Finite-Time response of the UDW Particle Detector in dS Space-time against the curvature of  space-time ($k$) for massless scalar fields. (From left to right) 1st, 2nd and 3rd columns show the plots for different values of $n$, with  $n=1$, $n=2$ and $n=3$. Each row (from top to bottom) shows the variation of the response function with changing $\Omega$ ($a=3, \T=3$), changing $a$ ($\T=3, \Omega=1$),  and changing $\T$ ($a=3, \Omega=1$) respectively.}
\end{figure}
Finally, to calculate the finite-time Unruh-DeWitt detector response function for dS spacetime, we plug in the 2n-point correlator $G_{{\rm dS}_D}^{(n)}$ from eq. (\ref{eqn:2nptfnmvrind}) to eq. \eqref{eqn:detrate} which gives,
\begin{align}\label{response1}
 \mathcal{F}^{(n)}_{{\rm dS}_D}(\Omega, \mathcal{T}) &= \frac{\pi\mathcal{T}^3}{4}\int^\infty_{-\infty}\dd{(\Delta \tau)}\frac{{G^{(n)}_{{\rm dS}_{D}}}(\Delta\tau)}{\Delta\tau^2 + \mathcal{T}^2}e^{-i\Omega\Delta\tau} \nonumber\\
&= \frac{\pi \T^3 n!}{4}\left(\frac{\Gamma(D/2-1)}{(4\pi)^{D/2}}\right)^n\left(\frac{\omega}{i}\right)^{n(D-2)}  \int^\infty_{-\infty}\dd{(\Delta \tau)}\frac{e^{-i\Omega\Delta\tau} }{\Delta\tau^2 + \mathcal{T}^2}\frac{1}{\sinh^{n(D-2)} (\frac{\omega\Delta\tau}{2}-i\epsilon)}\nonumber \\
&= \frac{\pi \T^3 n!}{4}\left(\frac{\Gamma(D/2-1)}{(4\pi)^{D/2}}\right)^n\left(\frac{\omega}{i}\right)^{n(D-2)} \left(\frac{\omega}{2}\right) \underbrace{\int^\infty_{-\infty}\dd{\rho}\frac{e^{-2i\Omega\rho/\omega} }{\rho^2 + \left(\omega\T/2\right)^2}\frac{1}{\sinh^{n(D-2)} (\rho-i\epsilon)}}_{\mathcal{F}_{D,n}}\nonumber \\
&= \frac{\pi \T^3 n!}{4}\left(\frac{\Gamma(D/2-1)}{(4\pi)^{D/2}}\right)^n\left(\frac{\omega}{i}\right)^{n(D-2)}\left(\frac{\omega}{2}\right)\cdot \mathcal{F}_{D,n}
\end{align}

where, 
\begin{equation}
    \mathcal{F}_{D,n} = \int^\infty_{-\infty}\dd{\rho}\frac{e^{-2i\Omega\rho/\omega} }{\rho^2 + \left(\omega\T/2\right)^2}\frac{1}{\sinh^{n(D-2)} (\rho-i\epsilon)}
\end{equation}
Finally, we evaluate the finite time response function $  \mathcal{F}^{(n)}_{{\rm dS}_D}$ numerically from \eqref{response1}  and plot it with respect to  energy gap,
acceleration,
and curvature as before.
Just like the case of AdS, we can also see that as the energy gap increases the response function decreases (Fig. 5). In the first row of figure 5 we can see if the value of acceleration is higher the response function.
\section{Finite time response of UDW detector: Dirac fields. }
\subsection{Dirac fields in AdS}
In the similar fashion we can analyse the response function for fermions in AdS spacetime minimally coupled to 
background gravity. The fermionic matter field 
action is-
\begin{eqnarray}
    \mathcal{S}_{0}=\int d^Dx\sqrt{|g|}
    \bar\Psi
    i\slashed{D} \Psi.\label{ld}
\end{eqnarray}

We can consider usual interaction Hamiltonian\cite{ottoEngine},
	\begin{equation}
	\mathcal{H}_\text{Int}=\lambda\;
	\chi_\mathcal{T}(\tau)m(\tau)\;\mathcal{O}_{\Psi}[x(\tau)]\;,\label{xchu}
	\end{equation}
Here, the operator 
${O}_{\Psi}[x(\tau)]$
is the normal ordered bispinor,
\begin{eqnarray}
{O}_{\Psi}[x(\tau)]=:\bar{\Psi}[x(\tau)]\Psi[x(\tau)]:\label{xcha}
\end{eqnarray}
Using  Lorentzian switching function eq. \eqref{33} 
the response function for interaction Hamiltonian \eqref{xchu} takes the following form,
\begin{equation}
 \mathcal{F}^{(n)}(\Omega, \mathcal{T}) = \frac{\pi\mathcal{T}^3}{4}\int^\infty_{-\infty}\dd{(\Delta \tau)}\frac{{S^{(4)}_{\rm_{D}}}(\Delta\tau)}{\Delta\tau^2 + \mathcal{T}^2}e^{-i\Omega\Delta\tau}  \label{ycha}.
\end{equation}
Here, the four point function 
$S^{(4)}_{\rm_{D}}$,
\begin{eqnarray}
     S^{(4)}_{D} (x(\tau),x(\tau')) &=& \bra{0}:\big( \overline{\Psi}_a(x(\tau))) \Psi_a(x(\tau)) \big)::\big( \overline{\Psi}_b(x(\tau'))) \Psi_b(x(\tau'))  \big):\ket{0}\nonumber\\ &=& \text{Tr}[S^{+}(x,x')S^{-}(x',x)]. \label{traces}
 \end{eqnarray}
As discussed in \cite{ahmed2021accelerated} for fermions in AdS spacetime,
  \begin{eqnarray}
        S_D^{(2)}(\Delta \tau)=N \frac{(\Gamma(D/2))^2}{\Gamma(D-1)}G_{AdS_{2D}}(\Delta \tau) \label{mmmmm}
    \end{eqnarray}
    From eq. \eqref{chhapu} we can then conclude that detector response function for fermions can be related to response function for the scalars,
    \begin{eqnarray}
        \mathcal{J}_{AdS_D}(\Delta \tau)= N \frac{(\Gamma(D/2))^2}{\Gamma(D-1)}
              \mathcal{F}^{(1)}_{\rm AdS_{2D}}(\Delta \tau). \label{pqrst}
\end{eqnarray}
In the next section we work on fermions in dS spacetime and we work out relations similar to eq. \eqref{mmmmm}. Therefore, we further demonstrate  that eq. \eqref{pqrst}
holds for maximally symmetric spacetime. The proof is similar to the case of AdS but we explicitly demonstrate it for the sake of completeness. 
  \begin{figure}[!h]
\centering 
{\large \textbf{\qquad \qquad \underline{\underline{$S_{{\rm AdS}_4}(\Omega, \T)$ }}}}\\
\hfill 
\includegraphics[scale=0.4]{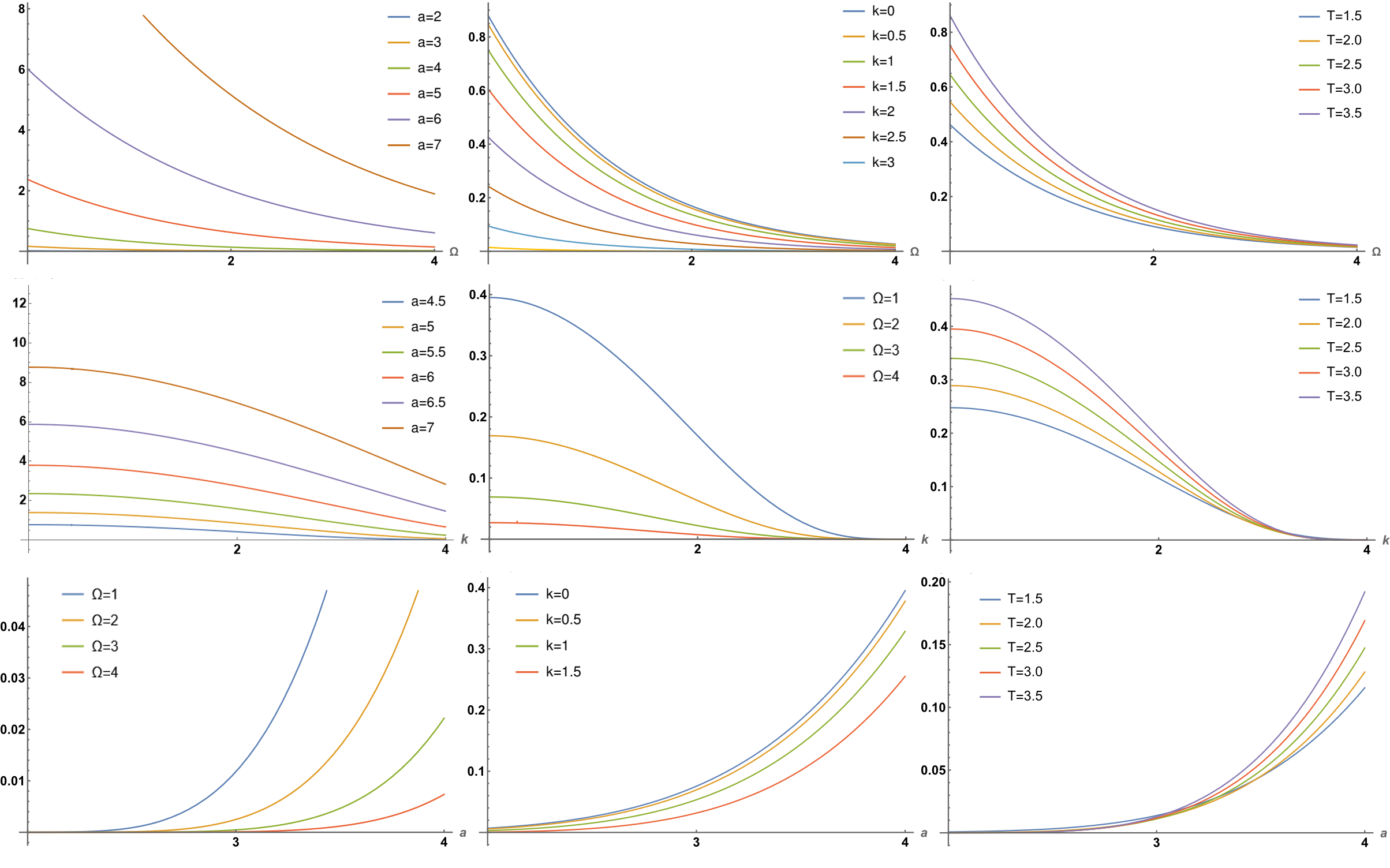}
\hfill
\caption{\label{fig:ads_fermion_all} Plot of the Finite-Time response of the UDW Particle Detector in AdS Space-time coupled to a fermionic Field. Each row (from top to bottom) shows the plot of the response function against $\Omega$ (varying $a$[$\T=3, k=1$], varying $k$[$\T=3, a=4$] and varying $\T$[$a=4, k=1$]), against $k$ (varying $a$[$\Omega=1, \T=3$], varying $\Omega$[$a=4, \T=3$] and varying $\T$[$a=4, \Omega=1$]), and against $a$ (varying $\Omega$[$k=2, \T=3$], varying $k$[$\Omega=1, \T=3$] and varying $\T$[$k=2, \Omega=1$]) respectively.}
\end{figure}

\subsection{Dirac fields in dS}

In order to study Dirac field in de Sitter spacetime we choose a local Lorentz frame (vielbein) which is defined as, $    e^a_\mu={\delta^a_\mu}/{(H\tau})$ such that $    g_{\mu\nu}=e^a_\mu e^b_\nu \eta_{ab}$ mimics the de-Sitter metric. Here Latin letters
\(a, b\) corresponds to local orthonormal flat  coordinates and Greek letters
\(\mu, \nu\) signifies the de Sitter coordinates. Both of them takes value from
\(0\) to \(D-1\). Also $\eta_{ab} = \text{diag}(+1,-1, ~\ldots{} ~,
-1) $ is the local flat metric. The vielbeins follow the usual orthonormal relations. Now, the curved space \(\Gamma\) matrices and the
covariant derivatives are defined as, \begin{eqnarray}
   && \Gamma^\mu= e_{a}^\mu\gamma^a,\nonumber\\
&& D_\mu=    \partial_{\mu} + \frac{1}{2}\omega^{bc}_{\mu} \Omega_{bc}, 
\end{eqnarray} 
where  \(\gamma_a\) are gamma matrices in  flat spacetime. And
 commutator between \(\gamma\) matrices are identified as
\(\Omega_{bc}\).
\begin{eqnarray}
    \Omega^{bc} = \frac{1}{4}[\gamma^b, \gamma^c]
\end{eqnarray} and the spin connections \(\omega^{bc}_{\mu}\) are noted as,
\begin{eqnarray}
    \omega^{ab}_{\mu}=e^{a \lambda }\big(\partial_\mu e^{ b}_{\lambda}-\bigg\{
    \begin{array}{cc}
    \alpha\\
    {\mu \ \ \lambda}
    \end{array}\bigg\}e^{b}_{\alpha}\big)\label{spinConn}
\end{eqnarray} and $\bigg\{
    \begin{array}{cc}
    \alpha\\
    {\mu \ \ \lambda}
    \end{array}\bigg\}$ are the Christoffel
symbols related to dS spacetime metric eq. (1). Here \(\Gamma^{\mu}\)
and \(\gamma^a\) maintain the well-known Clifford algebra,
\begin{eqnarray}
    \{ \Gamma^{\mu}, \Gamma^{\nu} \} &=  2g^{\mu \nu} \mathbb{I}_{N \times N}  \nonumber\\
     \{\gamma^{b}, \gamma^{c} \} &= 2 \eta^{b c} \mathbb{I}_{N \times N}.  \label{diracMin}
\end{eqnarray} with, \begin{eqnarray}
N =
  \begin{cases}
    2^{\frac{D}{2}} & \text{$D$ is even} \\
    2^\frac{D-1}{2} & \text{$D$ is odd}. 
  \label{chandumama}  \end{cases}    
\end{eqnarray}
In dS spacetime,
the Dirac operator takes the  form,
\begin{eqnarray}
   \slashed{D}= \Gamma^{\mu}{D_{\mu}}  \equiv e^{\mu}_{a}\gamma^{a}\big( \partial_{\mu} + \frac{1}{2}\omega^{bc}_{\mu} \Omega_{bc} \big)  = k\eta\bigg( \gamma^a \partial_{a} - \frac{D-1}{2\eta} \gamma^0 \bigg) .
\end{eqnarray}
To derive this relation we used 
\begin{eqnarray}
    \bigg\{
    \begin{array}{cc}
    \alpha\\
    {\mu \ \ \lambda}
    \end{array}\bigg\} &=&\frac{1}{\eta}\big( g^{\alpha 0}g_{\mu \lambda} - \delta^{\alpha}_{\lambda}\delta^{0}_{\mu} - \delta^{\alpha}_{\mu}\delta^{0}_{\lambda}\big),\\
    \omega^{ab}_{\mu} &=& \frac{g^{\beta 0}}{\eta}\big( e^b_{\mu}e^{a}_{\beta}-e^{a}_{\mu}e^{b}_{\beta} \big).
\end{eqnarray}
In dS spacetime minimally coupled Dirac fermions with mass $m$ to background gravity will have the action,
\begin{eqnarray}
    \mathcal{S}_{0}=\int d^Dx\sqrt{|g|}
    (\overline{\Psi}
    i\slashed{D} \Psi-m\overline{\Psi}\Psi ).\label{ld}
\end{eqnarray}
We can split the $\Psi$ field in two parts, namely
positive and negative frequency modes.
 \begin{eqnarray}
     \Psi(x)=\Psi^{+}(x)+\Psi^{-}(x).
 \end{eqnarray}
These are the solution of massive Dirac equation derived from equation (\ref{ld})
\begin{equation}
i\slashed{D}\Psi -m\Psi =0\ .\label{Direq}
\end{equation}
We will first look into the positive energy mode solutions $\psi^{(+)}$ of (\ref{Direq}). These solutions are proportional to 
\(e^{i\mathbf{px}}\), where
\(\mathbf{x}=(x^{1},\ldots ,x^{D-1})\) and
\(\mathbf{p}=(p_{1},\ldots ,p_{D-1})\), \(\mathbf{px}=p_{l}x^{l}\), and
the summation runs over \(l=1,\ldots ,D-1\).
We now decompose the positive energy modes into upper and lower components, \begin{equation}
\psi^{(+)} =\left[
\begin{array}{c}
\psi _{+}(\eta) \\
\psi _{-}(\eta)%
\end{array}%
\right] e^{i\mathbf{px}}. \label{decomp1}
\end{equation}
This can be done by using the following explicit definition Gamma matrix representation,
\begin{equation}
\gamma ^{0}=\left[
\begin{array}{cc}
\mathbb{I}_{(N/2)\times(N/2)} & \mathbf{0}_{(N/2)\times(N/2)} \\
\mathbf{0}_{(N/2)\times(N/2)} & -\mathbb{I}_{(N/2)\times(N/2)}%
\end{array}
\right] ,\;\gamma^{a}=\left[
\begin{array}{cc}
 \mathbf{0}_{(N/2)\times(N/2)} & \sigma ^{a} \\
-\sigma ^{a} & \mathbf{0}_{(N/2)\times(N/2)}
\end{array}
\right] ,  \label{gamflat}
\end{equation} 
with \(a=1,\ldots ,D-1\). 
The definitions of \(\mathbb{I}_{(N/2)\times(N/2)}\) and 
\( \mathbf{0}_{(N/2)\times(N/2)}\)
can be found in \cite{ahmed2021accelerated}.
\begin{eqnarray}
    \sigma ^{a}\sigma ^{b}+\sigma ^{b}\sigma^{a} &=& 2\delta^{ab} \mathbb{I}_{(N/2)\times(N/2)}, \nonumber \\
    \sigma^{a \dagger} &=& \sigma^{a}. \nonumber
\end{eqnarray}
The Dirac equation is then reduced to subsequent form for positive energy modes,
\begin{equation}
\left( \partial_{0}-\frac{D-1}{2\eta}\pm \frac{im}{k\eta} \psi _{\pm } \right) -ip_{l}\sigma^{l} \psi _{\mp }=0.
\label{psipm}
\end{equation}
Using equation (\ref{psipm}) we can deduce two different second order differential equations
for the upper and lower components: 
\begin{equation}
\left( \eta^{2}\partial_{0}^{2}-(D-1)\eta\partial _{0}+p^{2}
\eta^{2}+\frac{(D-1)^{2}}{4}+\frac{D-1}{2}+\frac{m^{2}}{H^{2}}\mp \frac{im}{k}\right) \psi _{\pm }=0.
\label{Eqpsipm}
\end{equation} Making the
following
substitution, \begin{equation}
\psi _{\pm }(\eta)=\eta^{D/2}\chi _{\pm }(\eta),  \label{xipm}
\end{equation} equation (\ref{Eqpsipm}) is reduced to the following form, \begin{equation}
\left( \eta^{2}\partial_{0}^{2}+\eta\partial _{0}+(p\eta)^{2}-\left(\frac{im}{k} \pm \frac{1}{2}\right)^2\right) \chi _{\pm }=0.  \label{Eqxipm}
\end{equation}
Now we write the solutions of (\ref{Eqxipm}) as
\begin{eqnarray}
    \chi_{\pm}(\eta) &=& C_{\pm}H^{(2)}_{\frac{im}{k} \pm \frac{1}{2}}(p\eta) \label{chichi}
\end{eqnarray}
where $H^{(2)}_{\nu}(x)$ are Hankel function of second kind of order $\nu$. We only considered Hankel function of second kind as the solution for equation (\ref{Eqxipm}) because for positive energy solution in Bunch-Davies vacuum we demand $\psi^{(+)} \propto e^{-ip\eta}$ \cite{Collins} . The coefficients $C_+$ and $C_-$ are not independent of each other. We can find the relationship $C_- = -i p_b \sigma^b C_+ /p$ by inserting the solution matrices (\ref{chichi}) and (\ref{xipm}) into the equation (\ref{psipm}). Moreover, we require additional quantum numbers apart from \(\mathbf{p}\) to specify all the solutions. In order to do that we need to fix the spinor $C_{+}$. Here we take orthonormal basis for spinors by choosing 
\(C_{+}=C_{\beta }^{(+)}w^{(\sigma )}\), where \(C_{\beta }^{(+)}\) is a normalization constant and \(w^{(\sigma )}\), \(\sigma =\)
\(1,\ldots ,N/2\), are one-column matrices of \(N/2\) rows, with elements \(w_{l}^{(\sigma )}=\delta _{l\sigma }\). Combining with the negative energy solutions this set $\beta = (\mathbf{p}, \sigma)$ will form a complete set of quantum numbers. As a result, the positive-energy mode functions for Bunch-Davies vacuum takes the following form,
\begin{equation}
\psi _{\beta }^{(+)}=C_{\beta }^{(+)}\eta^{{D}/{2}}e^{i\mathbf{px}}\left[
\begin{array}{c}
 w^{(\sigma )} {k}^{(2)}_{\frac{im}{k} + \frac{1}{2}}( p\eta) \\
-\frac{ip_{b}\sigma^{b}}{p} w^{(\sigma )} {k}^{(2)}_{\frac{im}{k} - \frac{1}{2}}( p\eta)
\end{array}
\right] .  \label{psi+}
\end{equation}

The coefficient \(C_{\beta }^{(+)}\) in (\ref{psi+}) is set on from the normalization condition (using inner product defined over constant time hypersurface) \cite{sahalect}
\begin{equation}
\langle \psi^{(+)}_{\beta}, \psi^{(+)}_{\beta'} \rangle = \int d^{D-1}x\,\sqrt{\frac{|g|}{g^{00}}}\psi _{\beta
}^{(+)\dagger}\psi _{\beta ^{\prime }}^{(+)}=\delta (\mathbf{p-p}^{\prime })\delta
_{\sigma \sigma ^{\prime }}.  \label{NormCond}
\end{equation}   
After evaluating the inner product we find the normalization constant
\begin{equation}
 C_{\beta }^{(+)} =\frac{\sqrt{p} k^{\frac{D-1}{2}}e^{-i\phi/2}}{\sqrt{8 (2\pi)^{D-2}}}e^{\frac{m\pi}{2k}}.  \label{Cbet}
\end{equation} 
where \(\phi\) represents an arbitrary phase. The negative-energy mode functions \(\phi^{(-)}\) can be imposing the condition that \(\phi^{(-)} \propto e^{-i\mathbf{px}+ip\eta}\). By following the same procedure illustrated above we obtain the negative energy solution: 
\begin{equation}
\psi _{\beta }^{(-)}= \frac{\sqrt{p} k^{\frac{D-1}{2}}e^{i\phi/2}}{\sqrt{8 (2\pi)^{D-2}}}e^{-\frac{m\pi}{2k}} \eta^{{D}/{2}}e^{-i\mathbf{px}}\left[
\begin{array}{c}
 w^{(\sigma )} {H}^{(1)}_{\frac{im}{k} + \frac{1}{2}}( p\eta) \\
 \frac{ip_{b}\sigma^{b}}{p} w^{(\sigma )} {H}^{(1)}_{\frac{im}{k} - \frac{1}{2}}( p\eta)
\end{array}
\right] .  \label{psi-}
\end{equation} In the above equation $H^{(1)}_{\nu}(x)$ are Hankel function of first kind of order $\nu$. Therefore we have successfully evaluated  the complete set of solutions for the Dirac equation (\ref{Direq}). Now we can write artibitary spinor solution $\Psi(x)$ in the operator form.
\begin{eqnarray}
    \Psi(x) &=& \sum_{\sigma = 1}^{N/2} \int d\mathbf{p} \  \bigg( b_{\sigma}(\mathbf{p}) \psi^{(+)}_{\sigma}(\mathbf{p},  x) + d^{\dagger}_{\sigma}(\mathbf{p}) \psi^{(-)}_{\sigma}(\mathbf{p}, x)  \bigg) \\
    \overline {\Psi}(x) &=& \sum_{\sigma = 1}^{N/2} \int d\mathbf{p} \  \bigg( b^{\dagger}_{\sigma}(\mathbf{p}, \lambda) \overline{{\psi}_{\sigma}^{(+)}}(\mathbf{p},  x) + d_{\sigma}(\mathbf{p}) \overline{\psi_{\sigma}^{(-)}}(\mathbf{p}, x)  \bigg) ,
\end{eqnarray}
where 
\begin{eqnarray}
    b_{\sigma}(\mathbf{p}) \ket{0} &=& d_{\sigma}(\mathbf{p}) \ket{0} = 0,\\
    \overline{\psi} &=& \psi^{\dagger} \gamma^{0},\\
    \{b_{\sigma}(\mathbf{p}), b^{\dagger}_{\sigma'}(\mathbf{p'}) \} &=& \delta({\mathbf{p - p'}}) \delta_{\sigma \sigma'}, \\
    \{d_{\sigma}(\mathbf{p}), d^{\dagger}_{\sigma'}(\mathbf{p'}) \} &=& \delta({\mathbf{p - p'}})  \delta_{\sigma \sigma'}.
\end{eqnarray}
Now we can have an explicit form of the Wightman functions of the fermionic field,
 \begin{eqnarray}
     S^{+}(x,x') &=&  \bra{0} \Psi(x)\overline{\Psi}(x')\ket{0}= \sum_{\sigma }  \int d\mathbf{p}  \  \psi^{(+)}_{\sigma}(\mathbf{p}, x)    
    \overline{{\psi}_{\sigma}^{(+)}}(\mathbf{p},  x') \nonumber \\
    &=&  \sqrt{\frac{\eta'}{\eta}} \left(i\left(\slashed{D} + \frac{\Gamma^0}{2\eta} \right) + m \right) 
     \Big( \mathcal{P}^{+}G_{dS_D}(x,x',\frac{im}{k}-\frac{1}{2})  + \mathcal{P}^{-}G_{dS_D}(x,x',\frac{im}{k}+\frac{1}{2})
     \Big)\nonumber\\
     \label{Splussss},\\
      S^{-}(x,x') &=&  \bra{0} \overline{\Psi}(x')\Psi(x)\ket{0} =  \sum_{\sigma }  \int d\mathbf{p}  \  \psi^{(-)}_{\sigma}(\mathbf{p}, x)\overline{{\psi}_{\sigma}^{(-)}}(\mathbf{p}, x') \nonumber\\
      &=& - \sqrt{\frac{\eta'}{\eta}} \left(i\left(\slashed{D} + \frac{\Gamma^0}{2\eta} \right) + m \right) 
     \Big( \mathcal{P}^{+}G_{dS_D}(x',x,\frac{im}{k}-\frac{1}{2})  + \mathcal{P}^{-}G_{dS_D}(x',x,\frac{im}{k}+\frac{1}{2})
     \Big)\nonumber\\ \label{sminus}
 \end{eqnarray} 
where $P^{\pm}= (\mathbb{I}_{N\times N} \pm \gamma^0)/2$ and 
\begin{eqnarray}
    G_{dS_D}(x,x',\frac{im}{k} \pm \frac{1}{2}) = \int d\mathbf{p} \frac{{(\eta \eta')}^{\frac{D-1}{2}} H^{D-2}}{8(2\pi)^{D-2}}   e^{i\mathbf{p(x-x')}} H^{(2)}_{\frac{im}{k} \pm \frac{1}{2}}(p \eta) \  H^{(1)}_{\frac{m}{k} \pm \frac{1}{2}}(p\eta' ).
\end{eqnarray}
 Now the Wightman function for 
 massless fermions in dS background,
\begin{eqnarray}
S^\pm(x,x')  = \pm i\sqrt{\frac{\eta'}{\eta}} \left(\slashed{D} + \frac{\Gamma^0}{2\eta}  \right)   G_{dS_D}(x,x')  \label{SSxx}
\end{eqnarray} and $G_{dS_D}$ is as usual from equations (\ref{two}-\ref{nu}) ,
\begin{eqnarray}
    G_{dS_D}(x,x')=G_{dS_D}(x',x, \frac{1}{2}) = G_{dS_D}(x,x', -\frac{1}{2})= \frac{H^{D-2}\Gamma(D/2-1)}{2(2\pi)^{D/2}} v^{1-D/2} . \label{scalarGGG} 
\end{eqnarray}\\
From equation (\ref{SSxx}) we can further deduce that 
\begin{eqnarray}
    S^\pm(x,x') = \pm i \frac{H\eta_{ab}(x^a-x'^a)\gamma^b}{\sqrt{\eta \eta'}v}\left(\frac{D-2}{2}\right)G_{dS_D}(x,x')
\end{eqnarray}
The detector is moving in with constant linear acceleration $a$ following \eqref{eq12} as before.
In that case we know the detector response function of  fermions (per unit time)
 for interaction Lagrangian is  given by\cite{Weinberg72}, 
 \begin{equation}
    \mathcal{J}_{dS_D}=\int_{-\infty}^\infty d\Delta\tau e^{-iE\Delta \tau} S^{(2)}_{D} (\Delta\tau) \label{chhapu}
 \end{equation}
\begin{figure}[tbp]
\centering 
{\large \textbf{\qquad \qquad \underline{\underline{$S_{{\rm dS}_4}(\Omega, \T)$}}}}\\
\hfill 
\includegraphics[scale=0.4]{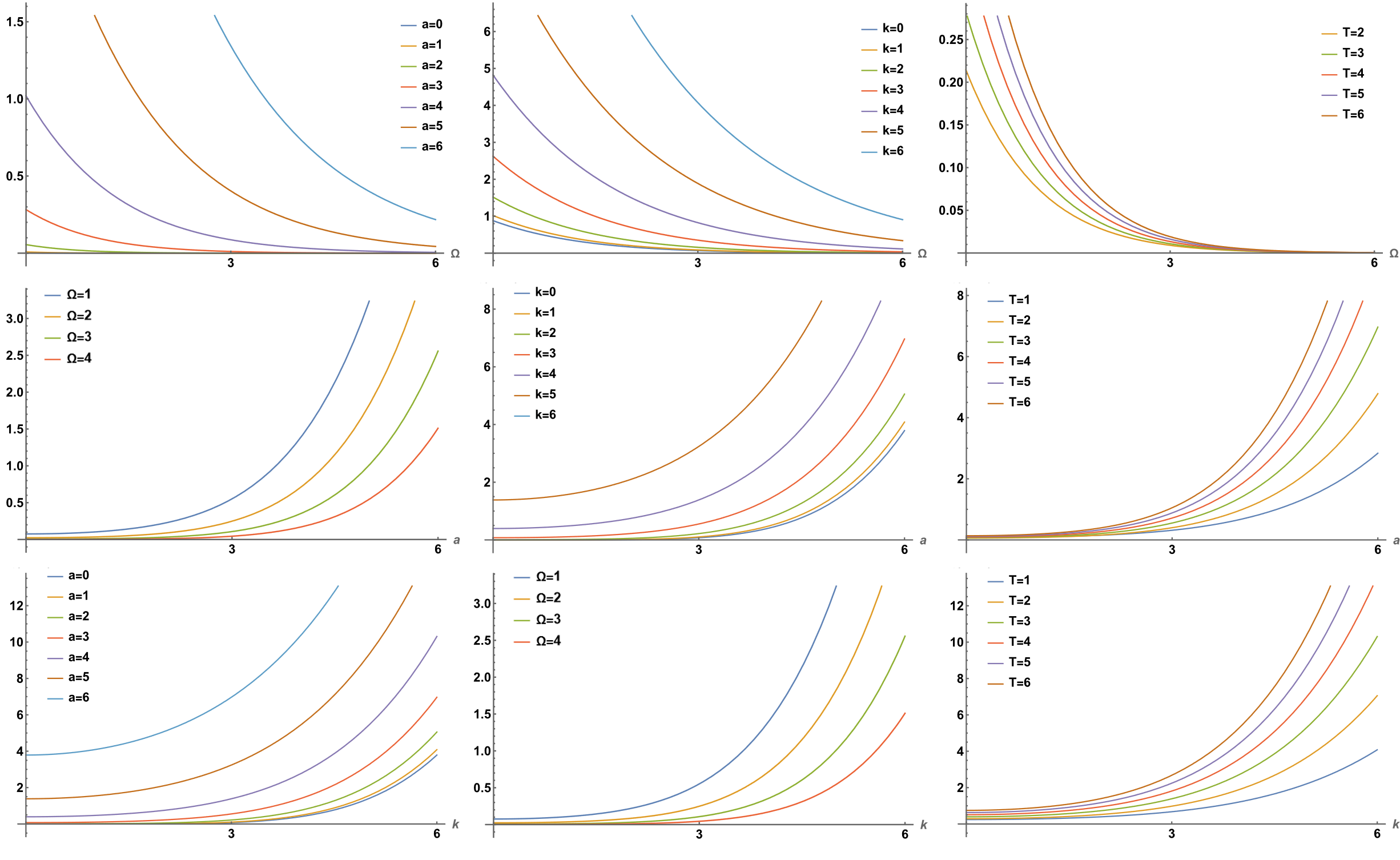}
\hfill
\caption{\label{fig:dS_ferm_response_all} Plot of the Finite-Time response of the UDW Particle Detector in dS Space-time coupled to a fermionic Field. Each row (from top to bottom) shows the plot of the response function against $\Omega$ (varying $a$[$\T=3, k=3$], $k$[$\T=3, a=3$], $\T$[$a=3, k=3$]), against $a$ (varying $\Omega$[$k=3, \T=3$], $k$[$\Omega=1, \T=3$], $\T$[$k=3, \Omega=1$]), and against $k$ (varying $a$[$\Omega=1, \T=3$], $\Omega$[$a=3, \T=3$], $\T$[$a=3, \Omega=1$]) respectively.}
\end{figure}

 where,
 \begin{eqnarray}
     S^{(2)}_{D} (x(\tau),x(\tau')) &=& \bra{0}:\big( \overline{\Psi}_a(x(\tau))) \Psi_a(x(\tau)) \big)::\big( \overline{\Psi}_b(x(\tau'))) \Psi_b(x(\tau'))  \big):\ket{0}\nonumber\\ &=& \text{Tr}[S^{+}(x,x')S^{-}(x',x)]\\
     &=& k^2 \frac{\eta_{ab}(x^a-x'^a)\eta_{cd}(x^c-x'^c)}{\eta\eta'\nu^2}\Tr[\gamma^b\gamma^d]\left(\frac{D-2}{2}\right)^2G_{{\rm dS}_D}(x,x')\,G_{{\rm dS}_D}(x',x)\nonumber\\
     &=&Nk^2\left(D/2-1\right)^2\frac{\eta_{ab}\eta_{cd}\eta^{bd}(x^a-x'^a)(x'^c-x^c)}{\eta\eta'\nu^2}\frac{H^{2D-4}\Gamma(D/2-1)^2}{4(2\pi)^D}\nu^{2-D}\nonumber\\
     &=& N \frac{\left[(D/2-1)\Gamma(D/2-1)\right]^2}{\Gamma(D-1)}\left(\frac{k^{2D-2}\Gamma(D-1)}{2(2\pi)^D}\right)\nu^{-D}\left(\frac{\eta_{ac}(x^a-x'^a)(x'^c-x^c)}{2\eta\eta'}\right)\nonumber\\
     &=& N\frac{\Gamma(D/2)^2}{\Gamma(D-1)}\left(\frac{k^{2D-2}\Gamma(D-1)}{2(2\pi)^D}\right)\nu^{1-D}\nonumber\\
     &=& N \frac{(\Gamma(D/2))^2}{\Gamma(D-1)}G_{{\rm dS}_{2D}}(x(\tau),x(\tau')). \label{issues}
 \end{eqnarray}
 is 4-points correlator of fermionic field. Here we are taking the trace over spinor index $a,b$, and, we have used the identity, 
 \begin{equation}
     \Tr[\gamma^b\gamma^d] = N\eta^{bd}
 \end{equation}
So the eq.\ref{issues} dictates that in any of the path $S_D^{(2)}(\Delta \tau)$ takes the following form,
\begin{equation}
    S_D^{(2)}(\Delta \tau) = N \frac{(\Gamma(D/2))^2}{\Gamma(D-1)}G_{{\rm dS}_{2D}}(\Delta \tau)
\end{equation}
From eq. \eqref{chhapu} we can then conclude that detector response function for fermions can be related to response function for the scalars,
\begin{eqnarray}
\mathcal{J}_{dS_D}= N \frac{(\Gamma(D/2))^2}{\Gamma(D-1)}
  \mathcal{F}^{(1)}_{\rm dS_{2D}}. \label{issue}
\end{eqnarray}
 Thus we have proved the following statement.
\\ \\
\normalsize	
\fbox{\begin{minipage}{45em}
 \large   The response function of an 
    UDW detector(with uniform linear acceleration)
      quadratically coupled to a massless Dirac field in (A)dS vacuum in $D \geq 2$ spacetime dimensions exactly equals to 
     the response function of a  UDW detector which is linearly coupled  to a massless scalar field in 
     $2D$ dimensional
     (A)dS vacuum   times  dimensional dependent    numeric factor. 
   Here, the  fermionic 
    field is minimally  coupled to background while the scalar field is
    conformally 
    coupled
    to the
background.
\end{minipage}}\\
    
     \large
Upon establishing the above statement for
fermionic response in maximally symmetric spacetime we plot the response function in figure 8 and 9 with respect to different variables such as energy gap $\Omega$, acceleration $a$ and curvature $k$.
We can also see the similar pattern in (A)dS fermionic response function
that the response increases with increasing acceleration $a$ but decreases as with increasing energy gap $\Omega$.  The response decreases with the increment curvature $k$ in AdS and the opposite pattern is noticed in dS background. 
Also because of the fact 4 dimensional fermionic response function is related to higher (eight) dimensional  scalar response function we find out, accelerated UDW detectors respond 
better when coupled to fermionic fields compared to bosonic fields.
\section{Huygen's principle, detector and Unruh radiation}
Huygen's principle is a well studied phenomenon specially in quantum field theory. It is a natural question to ponder whether the
accelerated detectors  observing
the thermal radiation 
maintains the Huygen's principle\cite{Ooguri:1985nv}.
The observed radiation from massless  scalars by UDW detectors
do not maintain the Huygen's principle in flat spacetime in three (odd) dimensions\cite{Ooguri:1985nv}.
However this statement is well
understood for accelerated UDW detectors in flat spacetime with linear coupling. But
in this section we discuss  the status of Huygen's principle for scalar theories where accelerated UDW detectors moving 
in the maximally symmetric curved spacetime with non linear interaction coupling\eqref{eq:SCoup}\cite{Sriramkumar:2002nt}.
The 
Huygen's principle 
 has several different 
 equivalent definitions but
we can work on with the  following one
\cite{Yagdjian:2020kkb,paul2014huygens,Jonsson}-\\\\
i) The  theory
\textbf{maintains} the Huygen's principle if the  causal propagator $G_c$ has support only on the  lightcone.
\\
ii) The  theory \textbf{violates} the Huygen's principle if they are 
non vanishing elsewhere.\\\\
To understand better the state of Huygen's principle for the detected Unruh radiation we
need to first fix the coupling between the detector and the matter field \eqref{chapo}. For the usual linearly coupled ($n=1$) detector, the response function simple depends upon the Wightman function.
Concentrating on linear coupling, the causal propagator for conformally coupled scalar theory can defined as,
\begin{eqnarray}
   G^c(x,x')=W^{(2)}_D(x,x')-W^{(2)}_D(x',x)=\bra{0}[{\Phi}(x),{\Phi}(x')]\ket{0}\label{kraikoo}
\end{eqnarray}
\begin{figure}[tbp]
\centering 
\hfill 
\includegraphics[scale=0.4]{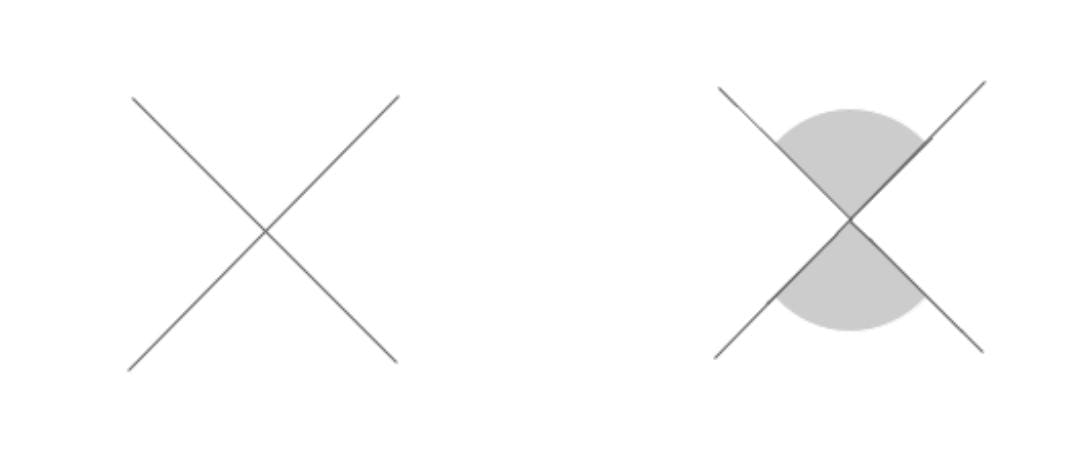}
\hfill
\caption{\label{fig:eldschapo} (A) Support of the propagators of a massless scalar in even dimensions for odd or even coupling associated with \eqref{eq:SCoup}. This also depicts
support of the propagators of a massless scalar in odd dimensions with even coupling. 
(B)
 Support of the propagators of a massless scalar in odd dimensions for odd  coupling associated with \eqref{eq:SCoup}
}
\end{figure}
In flat spacetime, the origin of obeying (or violating)
the Huygen's principle can be explained  for linearly coupled detector.
In case of the linear coupling the detector response function is nothing but the Fourier transform of the Wightman function $W_D^{(2)}(x,x')$,
which is proportional to $L^{1-\frac{d}{2}}$ for a
bosonic field. Here $L$ is the square distance between
the two points $x$ and $x'$. Now when the two point function is 
analytically continued
to complex $\tau$ there is a branch cut for timelike distance when we are in odd dimensions. This branch cut becomes a simple pole in even dimensions. 
Therefore when we compute the response function in odd dimensions using \eqref{sss} the linear detector finds a branch cut in the integral expression and therefore 
reports a Fermi-Dirac distribution. The support of the propagator of scalar fields for linear detector\cite{Ooguri:1985nv} can be exactly \eqref{chapo} analysed with figure 10A.
For even dimensions,
the support of $G_c$ is only on the lightcone while in odd dimension the support of $G_c$ is on the entire timelike region.
In the case of conformally coupled scalar fields over (A)dS spacetime,
we can follow the same argument as before. For example, the two point correlator  of conformally coupled scalars  in dS can be simply related to
flat space correlator 
$W^{(2)}_{M}(x,x')$ using the followings relation\cite{sahalect,marcos},
\begin{eqnarray}
    W^{(2)}_{dS}(x,x')
    =(k^2\eta^2)^{\frac{d-2}{4}}
    W^{(2)}_{M}(x,x') (k^2\eta'^2)^{\frac{d-2}{4}}\label{pablokusha}
\end{eqnarray}
Therefore the pole structure 
for conformally coupled theories are similar to those of flat spacetime. 
In even dimensional flat spacetime the causal correlator is given by\cite{Jonsson},
\begin{eqnarray}
  G_c=  [\Phi(t,\vec{x}),\Phi(t+\Delta t,\vec{x}+\Delta\vec{x})]=\frac{i}{4\pi\Delta\vec{x}}
  [\delta(\Delta \vec{t}+\Delta \vec {x})-
  \delta(\Delta \vec{t}-\Delta \vec {x})]\label{even}
\end{eqnarray}
 In similar fashion with odd 
 dimensional spacetime $D=d+1$\footnote{even dimensional space $d$.}, it is given by
\begin{eqnarray}
      G_c=  [\Phi(t,\vec{x}),\Phi(t+\Delta t,\vec{x}+\Delta\vec{x})]=\frac{\Gamma(\frac{d-1}{2})}{4\pi^{\frac{d+1}{2}}}\frac{1}{
      L^{\frac{d-1}{2}}}\label{odd}
\end{eqnarray}
We can see from eq. \eqref{even}
that in even dimensional Minkowski spacetime the support of $G_c$ is exactly on the lightcone while in odd dimensional spacetime the support of $G_c$ is also inside the lightcone. Exactly similar result will hold for conformally coupled scalar theory 
living on (A)dS background through 
\eqref{pablokusha}.
We now go ahead and generalize the results with
coupling $n\geq 1$.
The causal propagator can be written for any coupling $n$
in this fashion,
\begin{eqnarray}
    G^c(x,x')=W^{(2n)}(x,x')
-W^{(2n)}(x',x)\label{jjhh}
\end{eqnarray}
The $2n$ point correlators are related to two point correlators using \eqref{eqn:2nptfnmv}.
And thus the pole structure of $2n$ point correlator can be easily understood using 
this relation. In odd dimension the 
branch cut in Wightman function
results a branch cut in $2n$ point correlator for any odd coupling $n$. However when we choose the even coupling the branch cut  turns into simple pole for the $2n$ point correlator.
Huygen's principle for 
scalar fields are therefore obeyed
as well as no statistics inversion is noticed
by the Unruh radiation
in odd dimension only when we choose the even coupling. On the contrary the Huygen's principle is violated in odd dimensions with odd coupling and statistics inversion also happens.
 \\\\
In case of even coupling, the pole structure of the $2n$-point correlator function are also 
quite interesting. Through \eqref{eqn:2nptfnmv}
focusing in even dimensions,
the Wightman function has no branch cut in even dimensions. Therefore for any even coupling the $2n$-point correlators will not have any surprise branch cut in even dimensions. So, the Huygen's principle is going to be satisfied trivially. However, 
in odd dimensions there is a branch cut in Wightman function. 
But in the 
$2n$-point correlator,
when we use 
\eqref{eqn:2nptfnmv}
it immediately suggests the branch cut turns to a simple pole for any even $n$.
As a result for even $n$, the Unruh radiation of
scalars in flat space as well as in conformally coupled maximally symmetric scalar solutions the Huygen's principle is always maintained in any dimensions.
It is not possible to violate Huygen's principle if we choose coupling with even $n$.
The support of the 
scalar solutions are surprisingly always on the lightcone for even coupling. Figure $10 (A)$ accurately
depicts the status Huygen's Principle in scalar Unruh radiation with odd coupling and odd dimesnions, where the support is just on the light cone. Interesting to see this is the exact situation when the statistics inversion happens through \eqref{ss}. In odd dimensions
scalar theory propagator under consideration is anti-periodic in 
$\beta=\frac{2\pi}{\omega}$ when we choose odd coupling. In any other scenario the $2n$ point propagator is periodic in $\beta$ where the Hugen's principle is perfectly maintained in Unruh radiation. 
\begin{figure}[tbp]
\centering 
\hfill 
\includegraphics[scale=0.4]{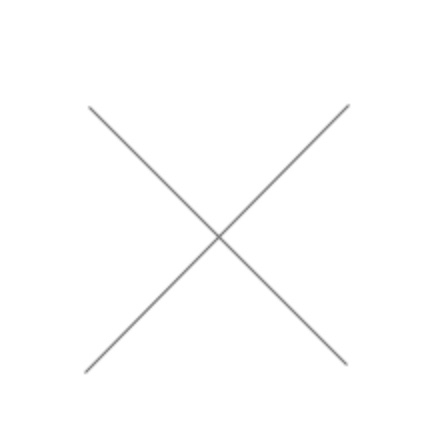}
\hfill
\caption{\label{fig:eldschapo}  
 Support of the four point propagators of a massless fermions in any dimensions for interaction Hamiltonian \eqref{xchu}. This is the clear indication that Huygen's principle is always maintained for fermions.
}
\end{figure}\\\\
Let us now focus our discussion on
the Huygen's principle for fermionic theory with interaction Hamiltonian \eqref{xcha}.
This is the most commonly used interaction Hamiltonian between fermionic theory and detector\cite{louko2016unruh}.
Now the definition of Huygen principle (written before eq. \eqref{kraikoo}) 
remains the same for fermionic theory as well but the definition of $G_c$ is given by Anti-Commuatator instead of commuatator algebra as in \eqref{kraikoo}\footnote{where the trace
over spin index is assumed},
\begin{eqnarray}
    G^c(x,x')=\bra{0}
\{\chi(x),\chi(x') \}\ket{0}
\end{eqnarray}
where,
\begin{eqnarray}
\chi(x)=:\overline{\Psi}_a(x) \Psi_a(x):
\end{eqnarray}
The study of support for
fermionic correlator on lightcone is a bit troublesome specially in the curved spacetime
because of the gamma matrices.
For the interaction Hamiltonian given by \eqref{xcha}, the detector response function is dependent upon the four point fermionic correlator as seen explicitly from
\eqref{ycha}. In ref.\cite{Ooguri:1985nv}, the author focused on the two point function of fermionic fields to understand the status of Huygen's principle of the Unruh radiation detected by the Unruh-DeWitt detectors instead of the four point function.
As a result, 
it was concluded in ref.\cite{Ooguri:1985nv} that
the Unruh radiation detected for fermions maintain Huygen's principle in odd dimensions while violate it in even dimensions.
Because the pole and branch cut structure of four point fermionic correlator is quite different to the two point correlator. In case of flat space\cite{louko2016unruh}, AdS\cite{ahmed2021accelerated} as well as in dS (\eqref{issue}), one can easily see that four point
 fermionic correlators
in $D$ dimensions is explicitly given by the 
two point scalar correlators in $2D$ dimensions. Therefore to understand the status of Huygen's principle 
for Unruh radiation of the fermionic theory \eqref{xchu} 
in odd or even $D$ dimension,
we can 
instead
think of
the 
massless scalars in $2D$ dimensions.
The scalar Wightman function when conformally coupled to the background (A)dS gravity solutions has no branch cut in even dimensions.
As a consequence,
the 4-point fermionic propagator, to which the detector is sensitive
always have support only on the light
cone.
It is quite surprising to see that scalar theory fails to maintain the Huygen principle in odd dimensions with usual linear coupling while
the fermionic theory always maintains the Huygen principle with the usual interaction Hamiltonian\eqref{xchu}.
This result is true for matter fields in flat spacetime as well as conformally coupled to maximally symmetric spacetimes.
The reason behind the conclusion being different to Huygen's principle of fermionic
Unruh radiation in ref.\cite{Ooguri:1985nv} 
is because they performed their analysis with two point function.
But for the interaction Hamiltonian
\eqref{xchu}\footnote{see 2.15b no eq. of \cite{Ooguri:1985nv}, where they use the same interaction Hamiltonian.},
one should actually
analyse the four point fermionic correlator. Because the detector response is dependent on four point fermionic correlator rather than the 
fermionic Wightman function (see \eqref{ycha}), the correlator under consderation maintains a  periodic condition with periodicity $\beta=\frac{2\pi}{\omega}$ and the Huygen's principle is always maintained   the irrespective of dimensionality of spacetime.
\section{Discussion and future works}
In this article we have completed the computation of finite time response of accelerated UDW detectors in maximally symmetric spacetime. The behaviour of the response function with different parameters are systemtically analysed in figure (2) - (9).  We also concluded the 
analysis for fermionic response function in maximally symmetric background 
(see the boxed statement after \eqref{issue}).
The result is quite powerful which also allows us to determine the status of Huygen's principle of the fermionic Unruh radiation detected by UDW detector moving in maximally symmetric spacetime.
It is quite intriguing to note that the Huygen's principle is always maintained in fermionic Unruh radiation which is minimally coupled to background as opposed to minimally coupled scalar\cite{Blasco:2015eya}.
We are currently working on to see if such results of fermionic response also holds when the UDW detectors are moving in other interesting curved spacetime solutions such as blackholes.
As an application of finite time response we would also like to construct 
Unruh Otto engines\cite{ottoEngine} with the help of UDW detectors moving in maximally symmetric backgrounds.
The variation of response function in dS and AdS space with respect to curvature makes the situation quite interesting and we are focusing on
explicit conditions to extract required conditions for completing Otto cycle to have positive work output.
The recent claim that with
entangled qubits one can 
build up more efficient\cite{Barman:2021igh,Kane:2021rhg}
Otto engine is quite exciting and we are exploring the possibility to generalise it with maximally symemtric spacetimes.
In our current manuscript, we have worked with scalar theory which is conformally coupled to background gravity but we are in a process of computing the finite time response function of UDW detectors for minimally coupled scalar theory\cite{bjorn}. 


\section{Appendix: Constant Acceleration Path in dS spacetime}\label{dS_acc_proof}
Here, we show that the path considered in eq. \ref{eq12} is a constant acceleration path. The components of the acceleration can be written as,
\begin{equation}
    a^{\mu} = \dv[2]{x^{\mu}}{\tau} + 2\Gamma^{\mu}_{\alpha\beta} \left( \dv{x^{\alpha}}{\tau} \right) \left( \dv{x^{\beta}}{\tau} \right)
\end{equation}
where, $\Gamma^{\mu}_{\alpha\beta}$ are Christoffel symbols of the first kind. Writing the components out explicitly for the path in eq. \ref{eq12}, we have, 
\begin{eqnarray}
    a^{(0)} &=& \dv[2]{\eta}{\tau} + 2 \Gamma^0_{\alpha\beta} \left( \dv{x^{\alpha}}{\tau} \right) \left( \dv{x^{\beta}}{\tau} \right)\nonumber \\ 
    &=& \dv[2]{\eta}{\tau} - \frac{1}{\tau}\left(\dv{\eta}{\tau}\right)^2 - \frac{1}{\tau}\left(\dv{x^1}{\tau}\right)^2\nonumber\\
    &=& \tau_0\omega^2e^{\omega\tau} - \frac{1}{\tau_0e^{\omega\tau}}\left(\tau_0\omega e^{\omega\tau}\right)^2 - \frac{1}{\tau_0e^{\omega\tau}}\left(\frac{a\tau_0}{\omega}\omega e^{\omega\tau}\right)^2\nonumber\\
    &=& -a^2\tau_0e^{\omega\tau} \\
    \nonumber\\
    a^{(1)} &=& \dv[2]{x^1}{\tau} + 2 \Gamma^1_{\alpha\beta} \left( \dv{x^{\alpha}}{\tau} \right) \left( \dv{x^{\beta}}{\tau} \right) \nonumber\\
    &=& \dv[2]{x^1}{\tau} - \frac{2}{\tau}\dv{x^1}{\tau}\dv{\eta}{\tau}\nonumber\\
    &=& \frac{a\tau_0}{\omega}\omega^2e^{\omega\tau} -\frac{2}{\tau_0e^{\omega\tau}}\left(\frac{a\tau_0}{\omega}\omega e^{\omega\tau}\right)\left(\tau_0\omega e^{\omega\tau}\right)\nonumber\\
    &=& -a\tau_0\omega e^{\omega\tau}\\
    \nonumber\\
    a^{(2)} &=&  a^{(3)} = \ ... \ = a^{(D-1)} = 0
\end{eqnarray}

So, the magnitude of the acceleration $\mathbf{a}$ becomes, 
\begin{align}
    |\mathbf{a}|^2 &= -a_{\mu}a^{\mu}\nonumber\\ 
    &=-g_{00}(a^0)^2 -g_{11}(a^1)^2\nonumber\\
    &=-\frac{1}{H^2\tau^2}a^4\tau_0^2e^{2\omega\tau} +\frac{1}{H^2\tau^2}a^2\tau_0^2\omega^2e^{2\omega\tau} \nonumber\\
    &=-\frac{1}{H^2\tau_0^2e^{2\omega\tau}}a^2\tau_0^2e^{2\omega\tau}(a^2-\omega^2) \nonumber\\
    &= a^2
\end{align}
Hence, $|\mathbf{a}| = a$ and the acceleration along this path is uniform.

\section{Acknowledgments}
MMF’s research is supported by NSERC and in part by the Delta Institute of Theoretical Physics. The authours would like to thank Sowmitra Das and Onirban Islam for the discussions.

\bibliographystyle{unsrtnat}  
\bibliography{template}  

\end{document}